\documentclass[pra,twocolumn,showpacs,unsortedaddress]{revtex4-1}

\usepackage{amsmath,amssymb}
\usepackage[pdftex]{graphicx}
\usepackage{nicefrac}
\usepackage{bbm}

\usepackage{color}

\newcommand{\ii}{\mathrm{i}}
\newcommand{\ad}{a^\dagger}

\let \Re \relax
\DeclareMathOperator{\Re}{Re}

\begin{document}

\title{Quantum phase transition in the Dicke model \\ with critical and non-critical entanglement}

\author{L. Bakemeier}
\affiliation{Institut f\"ur Physik, Ernst-Moritz-Arndt-Universit\"at, 17487 Greifswald, Germany}
\author{A. Alvermann}
\affiliation{Theory of Condensed Matter, Cavendish Laboratory, Cambridge CB3 0HE, United Kingdom}
\altaffiliation{Institut f\"ur Physik, Ernst-Moritz-Arndt-Universit\"at, 17487 Greifswald, Germany}
\email{alvermann@physik.uni-greifswald.de}
\author{H. Fehske}
\affiliation{Institut f\"ur Physik, Ernst-Moritz-Arndt-Universit\"at, 17487 Greifswald, Germany}

\begin{abstract}
We study the quantum phase transition of the Dicke model in the classical oscillator limit, where it occurs already for finite spin length.
In contrast to the classical spin limit,
for which spin-oscillator entanglement diverges at the transition,
entanglement in the classical oscillator limit remains small.
We derive the quantum phase transition with identical critical behavior in the two classical limits  
and explain the differences with respect to quantum fluctuations around the mean-field ground state through an effective model for the oscillator degrees of freedom.
With numerical data for the full quantum model we study convergence to the classical limits.
We contrast the classical oscillator limit with the dual limit of a high frequency oscillator,
where the spin degrees of freedom are described by the Lipkin-Meshkov-Glick model.
An alternative limit can be defined for the Rabi case of spin length one-half, in which spin frequency renormalization replaces the quantum phase transition.
\end{abstract}


\maketitle

\section{Introduction}

For a system of a single spin coupled to a quantum harmonic oscillator a quantum phase transition (QPT)~\cite{Sach00} can take place only in the classical limit of one of the two components, i.e. in the limit of infinite spin length or zero oscillator frequency.
Prior to the respective classical limit, the spin-oscillator system admits no phase transition since symmetry breaking states can always be combined in a linear superposition that restores the symmetry and reduces the energy further.
In the classical limit phase transitions become possible because different classical states have zero overlap, which circumvents the previous argument against symmetry breaking.

This type of QPT is realized in the  Dicke model~\cite{Di54} 
\begin{equation}\label{Dicke}
H  = \Delta J_z + \Omega a^\dagger a + \lambda (a^\dagger + a) J_x \;.
\end{equation}
It describes an ensemble of $2j$ two-level atoms with transition frequency $\Delta$
as a pseudo-spin of length $j$ (using spin operators $J_{x/z}$).
The atoms are coupled to a single cavity mode of the photon field with frequency $\Omega$ (using bosonic operators $a^{(\dagger)}$).
The Hamiltonian in Eq.~\eqref{Dicke} is invariant under the replacement $J_x \mapsto -J_x$, $a \mapsto -a$. This symmetry is broken in a phase transition,
and the spin expectation value $\langle J_x \rangle$ serves as the order parameter.

In the classical spin (CS) limit $j \to \infty$ the Dicke model features a thermodynamic phase transition from a 
high temperature state with $\langle J_x \rangle = \langle a \rangle = 0$ to a superradiant state with a finite cavity field
($\langle a\rangleÊ\ne 0$) and macroscopic atomic excitation ($\langle J_x \rangle \ne 0$) at low temperatures~\cite{HL73,WH73,GB76}.
The thermodynamic phase transition is complemented at zero temperature by a QPT 
 from the zero field to the superradiant state at a critical 
atom-field (i.e. spin-oscillator) coupling $\lambda_c$.

The driving mechanism behind the QPT is the critical behavior of a classical energy functional for the spin, which is obtained after integrating out the quantum-mechanical oscillator.
Strictly in the {$j=\infty$}--limit the ground state is a mean-field (MF) product state of a spin and  oscillator coherent state.
The order parameter $\langle J_x \rangle$
and the corresponding susceptibility $\chi$, which characterize the critical behavior,
converge to the classical results if the CS limit is approached from $j < \infty$.

Modifications of the classical picture arise from quantum corrections of order $1/j$ to the MF ground state~\cite{EB03,EB03b}. Spin and oscillator variances diverge and signal the breakdown of the classical limit in the vicinity of the QPT.
A characteristic feature is the criticality of spin-oscillator entanglement~\cite{LEB04,LEB05,VD06},
which is related to the vanishing excitation gap at the QPT 
and found for many different models in the CS limit~\cite{VDB07}.

In this paper we address a QPT in the different classical limit $\Omega\to 0$, the classical oscillator (CO) limit.
In contrast to the CS limit, a QPT transition occurs here already at finite spin length $j$.
The critical behavior is identical to the CS limit, since both limits realize the same MF transition.
Quantum corrections to the MF ground state are different.
In the CO limit spin fluctuations are suppressed because of the large spin frequency.
Therefore, the spin variance and the spin-oscillator entanglement remain small in the vicinity of the QPT. The entanglement entropy is bounded by $\ln 2$ independently of $j$.
The CS and CO limits thus give rise to QPTs with identical critical behavior that are distinguished through the criticality versus non-criticality of entanglement.

\begin{figure}
\centering
\includegraphics[width=0.5\linewidth]{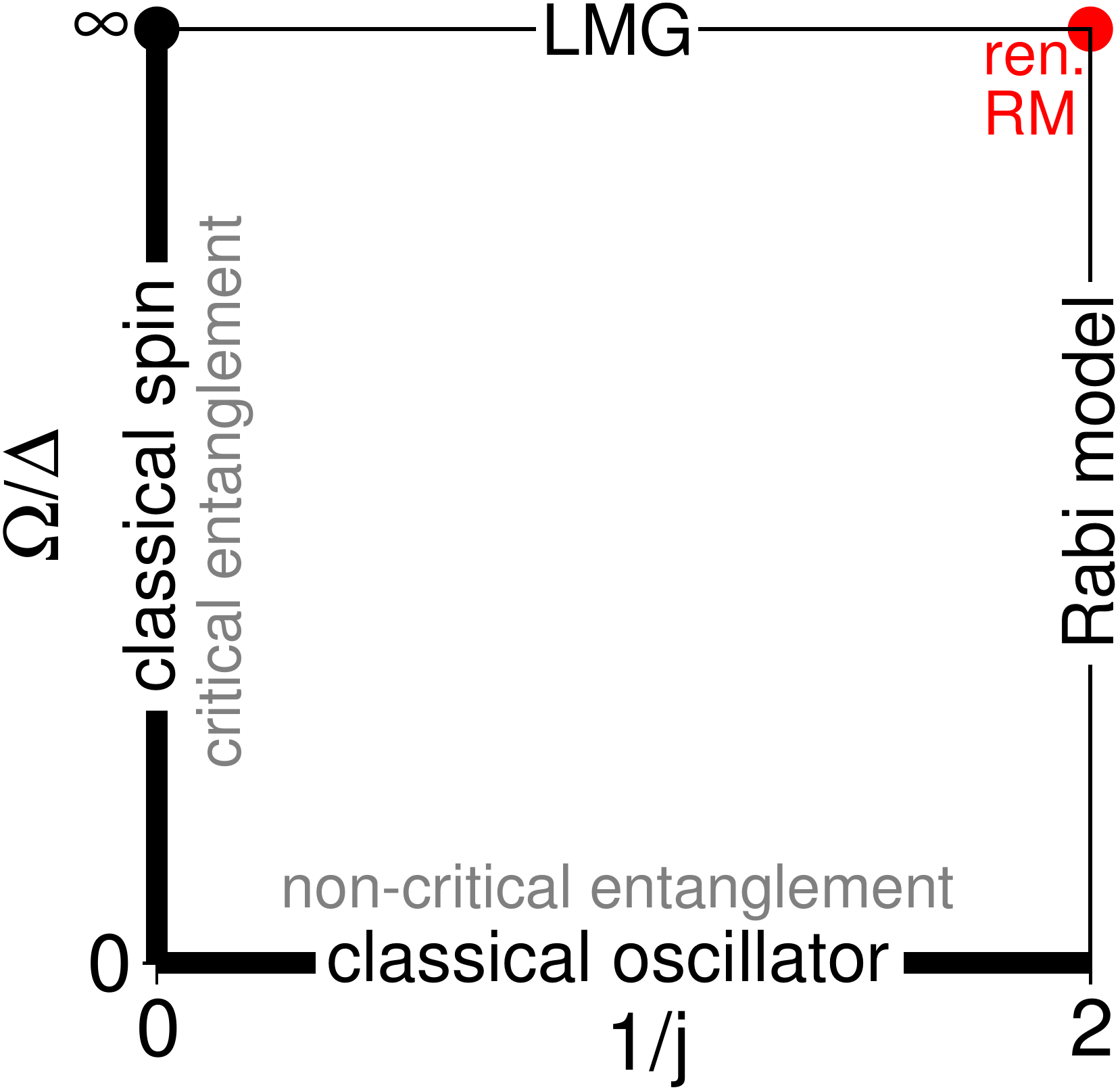}
\caption{(Color online) Diagram of the qualitative behavior of the Dicke model 
in dependence on the spin length $j$ and the spin/oscillator frequency ratio $\Omega/\Delta$.
The bold axes correspond to the QPT in the CS limit $1/j=0$ and the CO limit 
$\Omega/\Delta=0$.
The top edge of the square corresponds to the FO limit $\Omega/\Delta \to \infty$,
where the LMG model describes the spin. 
On the right side of the square we find the Rabi model ($j=1/2$),
with renormalization of the effective spin frequency in the limit $\Omega/\Delta \to \infty$ (upper right corner).
}
\label{fig:Dicke}
\end{figure}

An overview of the different limits in the Dicke model is given in Fig.~\ref{fig:Dicke}.
The paper is organized according to this diagram.
We first derive in Sec.~\ref{sec:QPT} the QPT in the CS and the CO limit from MF theory,
which becomes exact in the two limits.
Quantum corrections in the CO limit are discussed in Sec.~\ref{sec:QFLU} with an effective bosonic model for the oscillator degree of freedom, and contrasted with the behavior in the CS limit known from literature.
The (non-) criticality of entanglement is addressed in Sec.~\ref{sec:QENT}.
We complement the CO limit with the fast oscillator (FO) limit $\Omega/\Delta \to \infty$ in Sec.~\ref{sec:FO}.
Similar to the considerations for the CO limit,
the large oscillator frequency leads to the suppression of oscillator fluctuations.
One obtains the Lipkin-Meshkov-Glick (LMG) model for the spin degree of freedom,
but no QPT occurs unless we let again $j\to \infty$.
A related FO limit that is peculiar to the Rabi case $j=1/2$
leads to renormalization of the spin frequency instead of a QPT.
The appendices summarize the solution of the effective bosonic models for the CO and FO limit (App.~\ref{app:HOM}), and the definition of coherent states (App.~\ref{app:Coh})
and of the rotation invariant spin variance (App.~\ref{app:JVar}).

\section{Classical limits and the quantum phase transition} 
\label{sec:QPT}

As noted in the introduction, 
the Dicke Hamiltonian from Eq.~\eqref{Dicke}
is invariant under the symmetry transformation
\begin{equation}
\Pi = e^{i\pi N_E};\hspace*{0.2cm} N_E = \ad  a + J_z + j \;,
\end{equation}
which corresponds to the simultaneous replacement of $a \mapsto -a$ and $J_x \mapsto -J_x$.
We have $\Pi^{-1} H \Pi = H$ or $[H,\Pi]=0$,
such that the eigenstates of $H$ can be classified by the eigenvalues $\pm 1$ of the parity operator $\Pi$.
For positive $\Delta$, the ground state of the Dicke model has positive parity. 

The QPT breaks parity symmetry, with a finite order parameter  $\langle J_x \rangle \ne 0$ above a critical coupling.
To study the convergence to the classical limits it is useful to break the parity symmetry explicitly,
using the Hamiltonian
\begin{equation} \label{Heps}
 H_\epsilon = H - \epsilon J_x 
\end{equation}
that includes a symmetry breaking field $\epsilon J_x$.

\subsection{Mean-field theory of the QPT}

\begin{figure}
\includegraphics[width=0.41\linewidth]{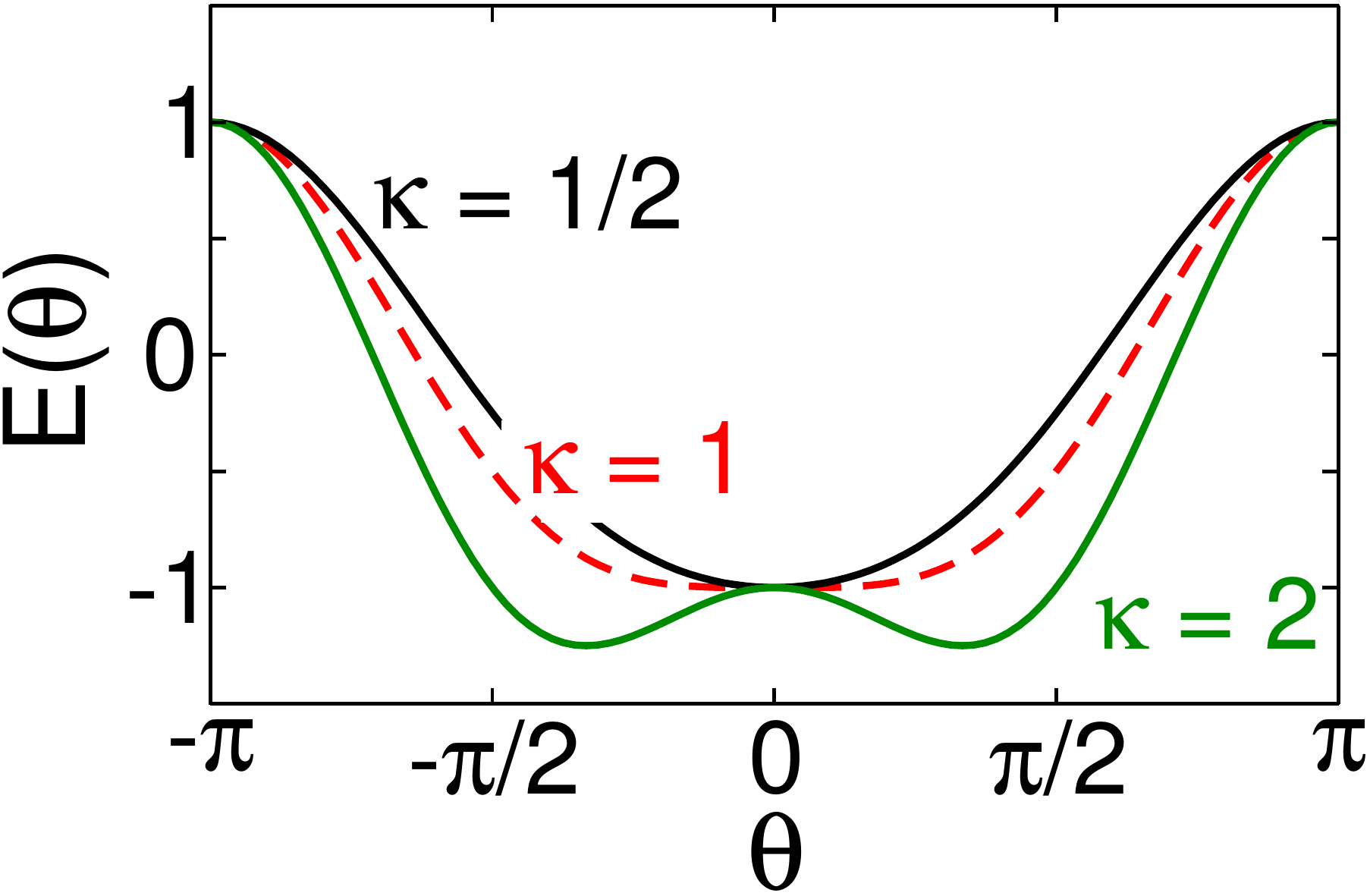} \hfill
\includegraphics[width=0.49\linewidth]{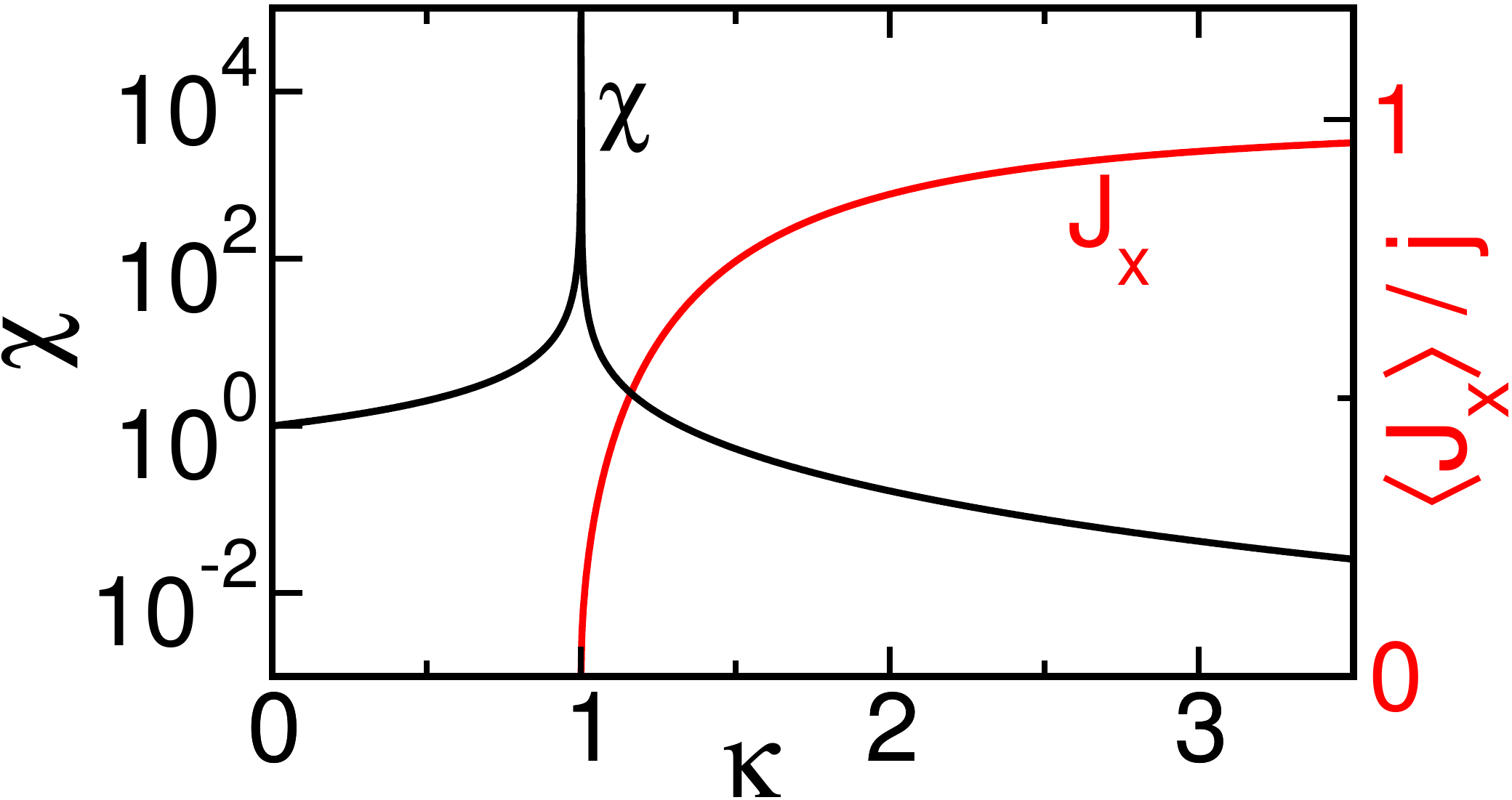} 
\caption{(Color online) Mean-field QPT in the Dicke model. 
Left panel: Energy functional $E(\theta)$ from Eq.~\eqref{ErgFunc} below ($\kappa < 1$),
above ($\kappa > 1$) and at ($\kappa = 1$, dashed curve) the QPT.
Right panel: Order parameter $\langle J_x \rangle$ (red) and susceptibility $\chi$ (black) of the mean-field QPT given in Eqs.~\eqref{MFJx},~\eqref{MFsus}.}
\label{fig:MF}
\end{figure}

Let us first discuss the QPT in the MF picture.
The MF ansatz for the ground state wave function
\begin{equation}\label{MFPsi}
 |\psi_\mathrm{MF}\rangle =  |\theta\rangle \otimes |\alpha\rangle
\end{equation}
is a product 
of a spin coherent state $|\theta\rangle$ and a boson coherent state $|\alpha\rangle$
(cf. App.~\ref{app:Coh}).
This state has energy
\begin{equation}\label{MFErg}
\begin{split}
  E (\theta,\alpha) &= \langle \psi_\mathrm{MF} | H | \psi_\mathrm{MF} \rangle \\
  &=  -j\Delta\cos\theta + \Omega\alpha^2 + 2 j\lambda\alpha\sin\theta \;.
 \end{split}
 \end{equation}
Minimization with respect to $\alpha$ results in 
\begin{equation}\label{MFAlpha}
 \alpha = - \frac{j \lambda}{\Omega} \sin \theta \;,
 \end{equation}
 which inserted into Eq.~\eqref{MFErg} gives the energy functional
 \begin{equation}\label{ErgFunc}
   E (\theta) = -j\Delta\left(\cos\theta + \frac{\kappa}{2}\sin^2\theta\right) \;.
\end{equation}
We here introduce the dimensionless coupling constant
\begin{equation}\label{LamTil}
 \kappa = \frac{2 j \lambda^2}{\Delta\Omega} \;.
\end{equation}
We will assume 
$\kappa \ge 0$, which corresponds to $\Delta > 0$.

The energy functional $E(\theta)$ describes a second-order MF transition at $\kappa=1$ (cf. Fig.~\ref{fig:MF}).
The minima of $E(\theta)$ are given by 
\begin{equation}\label{MFTheta}
  \theta = \begin{cases}
    0  \quad & \text{ if } \kappa < 1 \;, \\
    \pm \arccos \dfrac{1}{\kappa} & \text{ if } \kappa > 1 \;,
  \end{cases}
\end{equation}
which leads to the expression
\begin{equation}\label{MFJx}
 \langle J_x \rangle = j \sin \theta = 
 \begin{cases}
    0  \quad & \text{ if } \kappa < 1 \;, \\
    \pm j \sqrt{1-\dfrac{1}{\kappa^2}}  & \text{ if } \kappa > 1 
  \end{cases}
\end{equation}
for the order parameter $\langle J_x \rangle$.
We can also calculate the susceptibility $\chi$ using the Hamiltonian $H_\epsilon$ from Eq.~\eqref{Heps}, and find 
\begin{equation}\label{MFsus}
 \chi =  j \Delta \, \lim_{\epsilon \to 0} \frac{\partial \langle J_x \rangle}{\partial \epsilon}
 =  \begin{cases}
    \dfrac{1}{1-\kappa}  \quad & \text{ if } \kappa < 1 \;, \\[3ex]
     \dfrac{1}{\kappa (\kappa^2-1)} & \text{ if } \kappa > 1 \;.
  \end{cases}
\end{equation}
 
In contrast to the prediction of MF theory,
the argument given in the introduction shows
that a QPT cannot exist in the fully quantum-mechanical Dicke model for finite $j$, $\Omega/\Delta$.
The QPT only becomes possible if the two degenerate MF states for positive/negative $\theta, \alpha$ have zero overlap.
This is can be achieved either if $\langle \theta |{ - \theta }\rangle =0$ in the CS limit,
or if $\langle \alpha |{ - \alpha }\rangle =0$ in the CO limit.

\subsection{QPT in the classical spin limit}
In the CS limit $j \to \infty$ spin coherent states form an orthonormal basis of the spin Hilbert space
(see, e.g., Ref.~\cite{L73}).
Therefore, the ground state wave function has the form $|\psi_\mathrm{CS}\rangle = |\theta\rangle \otimes |\psi_\mathrm{bos}\rangle$, with
a real spin coherent state $|\theta\rangle$ as defined in Eq.~\eqref{app:theta} in App.~\ref{app:Coh}.
Note that for $\theta \ne 0$ the overlap $\langle \theta |{ - \theta }\rangle = \cos^{2j} \theta$
goes to zero for $j \to \infty$, which allows for the QPT.

The bosonic part $|\psi_\mathrm{bos}\rangle$ of the wave function,
which has to be determined through minimization of the energy $\langle \psi_\mathrm{CS} | H |\psi_\mathrm{CS} \rangle$, is the ground state of the effective bosonic Hamiltonian
\begin{equation}
H_\mathrm{osc}(\theta) = \Omega \ad a + \lambda j \sin \theta (a+\ad) \;, 
\label{eq:meanfieldOsci}
\end{equation}
which is parameterized by the classical spin angle $\theta$.

$H_\mathrm{osc}(\theta)$ is the Hamiltonian of an oscillator with a constant force $\propto \lambda j\sin\theta$.
The ground state of this Hamiltonian is a boson coherent state $|\psi_\mathrm{bos}\rangle = |\alpha\rangle$, with $\alpha$ given by Eq.~\eqref{MFAlpha}. We thus recover the MF wave function from Eq.~\eqref{MFPsi} in the CS limit, hence also the entire QPT.
 
 \subsection{QPT in the classical oscillator limit}\label{sec:QPTCO}
 
According to Eqs.~\eqref{MFAlpha},~\eqref{LamTil} 
the parameter $\alpha$ in the MF ground state in Eq.~\eqref{MFPsi} scales as $1/\sqrt{\Omega}$.
For $\Omega \to 0$ the overlap $\langle \alpha | {-\alpha}Ê\rangle = \exp(-2 \alpha^2)$
goes to zero for $\kappa>1$.
Because of this a QPT in the CO limit is possible independently of the spin length $j$.

Since the overlap of different real coherent states $|\alpha\rangle$ is zero in the CO limit,
the ground state wave function has the form $|\psi_\mathrm{CO}\rangle = |\psi_\mathrm{spin}\rangle \otimes |\alpha\rangle$.
The spin part $|\psi_\mathrm{spin}\rangle$ of the wave function is the ground state of the effective spin Hamiltonian
\begin{equation}
H_\mathrm{spin}(\alpha) = \Delta J_z + 2 \lambda \alpha J_x \;,
\label{eq:meanfieldSpin}
\end{equation}
parameterized by the classical oscillator displacement $\alpha$.

$H_\mathrm{spin}(\alpha)$ is the Hamiltonian of a spin in a magnetic field $\vec{B}=(2\lambda\alpha,0,\Delta)$,
with a coherent spin state $|\psi_\mathrm{spin}\rangle=|\theta\rangle$ as the ground state.
$\theta$ and $\alpha$ are related through Eq.~\eqref{MFAlpha}.
Again, we recover the MF wave function from Eq.~\eqref{MFPsi},
and therefore also the entire QPT in the CO limit.

Note that the argument for the CS and CO limit are dual to each other:
In the CS limit we first observe that the spin has to be in a coherent (``classical'') state and
deduce the oscillator coherent state from the particular effective Hamiltonian $H_\mathrm{osc}(\theta)$ for the quantum mechanical oscillator.
In the CO limit, we start from an oscillator coherent state and obtain the spin coherent state again only because of the particular form of the effective Hamiltonian $H_\mathrm{spin}(\alpha)$ for the quantum spin.

\subsection{Convergence to the QPT}

\begin{figure}
\includegraphics[width=0.48\linewidth]{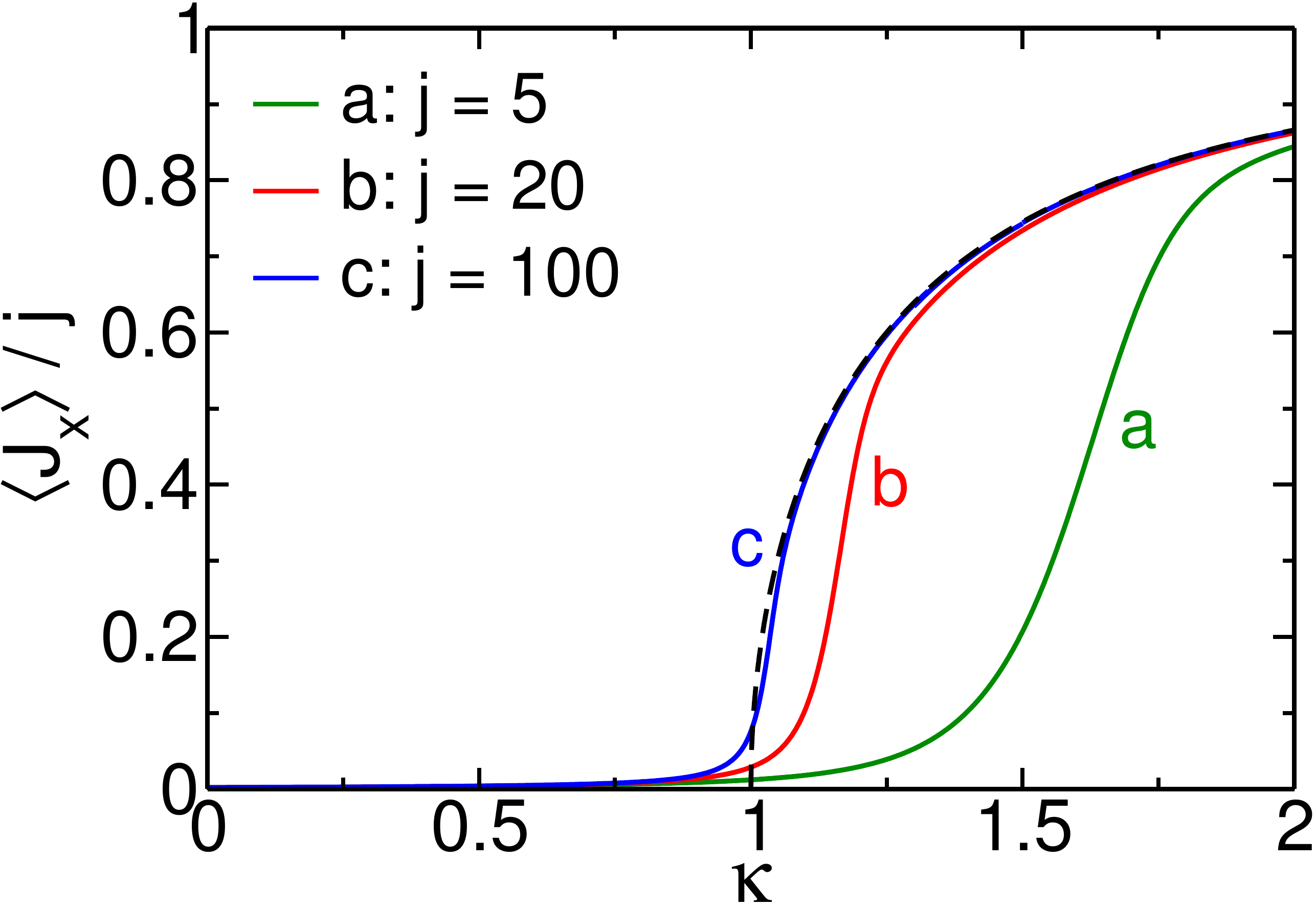} \hfill
\includegraphics[width=0.48\linewidth]{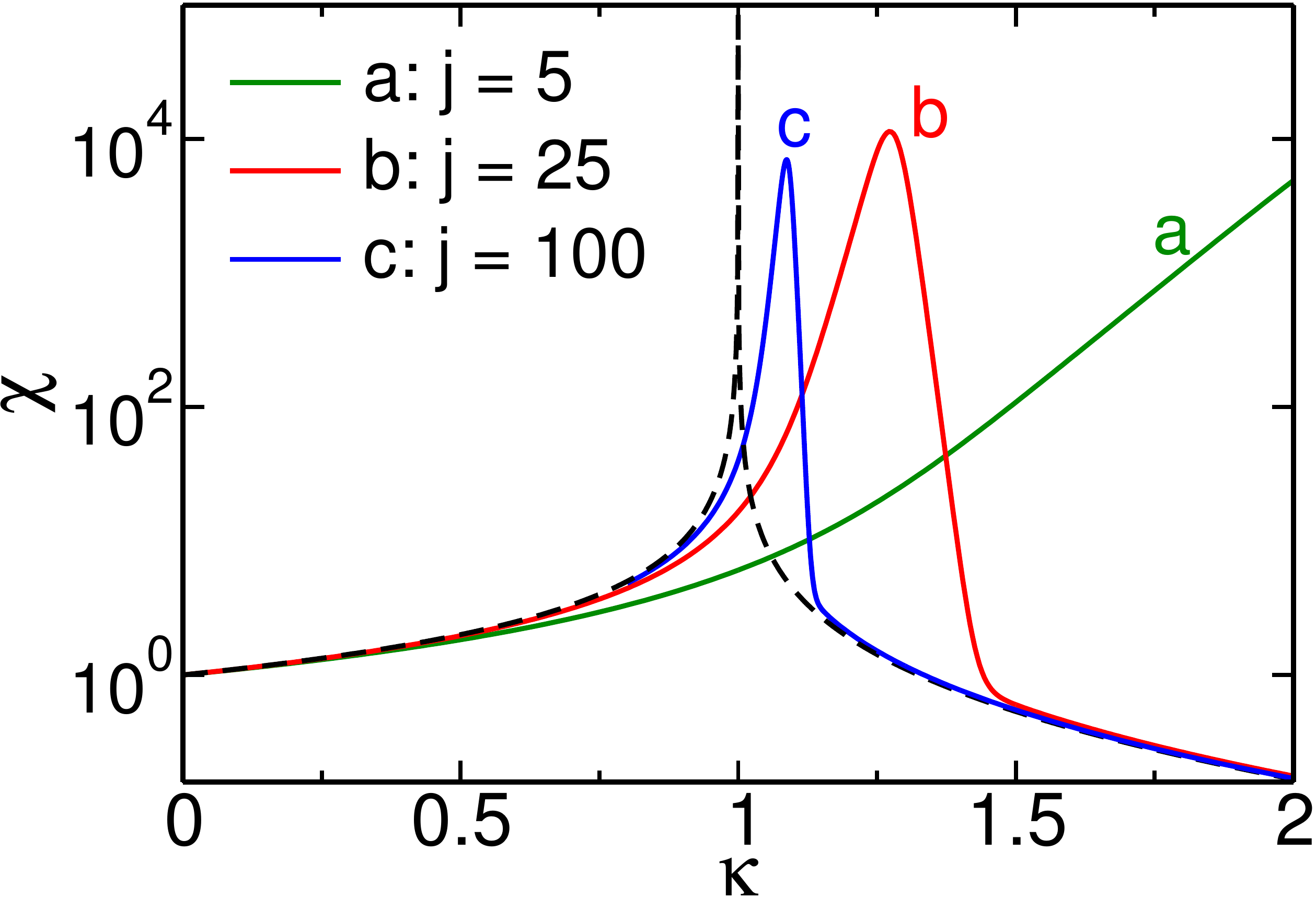} 
\caption{(Color online) QPT in the CS limit. 
Order parameter $\langle J_x \rangle$ (left panel) and
susceptibility $\chi$ (right panel) as a function of $\kappa$,
for various values of $j$ and fixed $\Omega/\Delta=1$.
The dashed curves give the MF result from Eqs.~\eqref{MFJx},~\eqref{MFsus}.}
\label{fig:CSQPT}
\end{figure}

\begin{figure}
\includegraphics[width=0.48\linewidth]{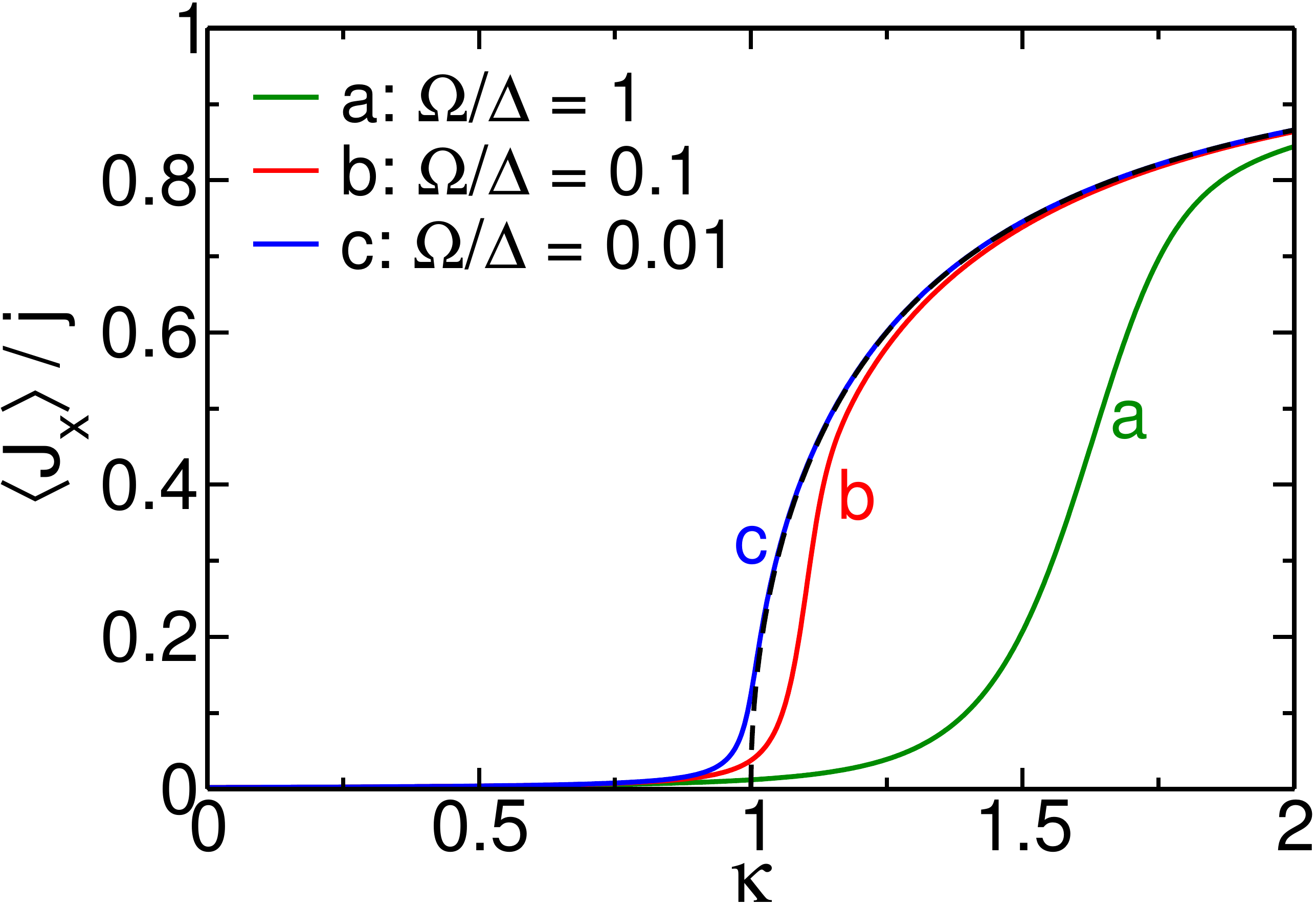} \hfill
\includegraphics[width=0.48\linewidth]{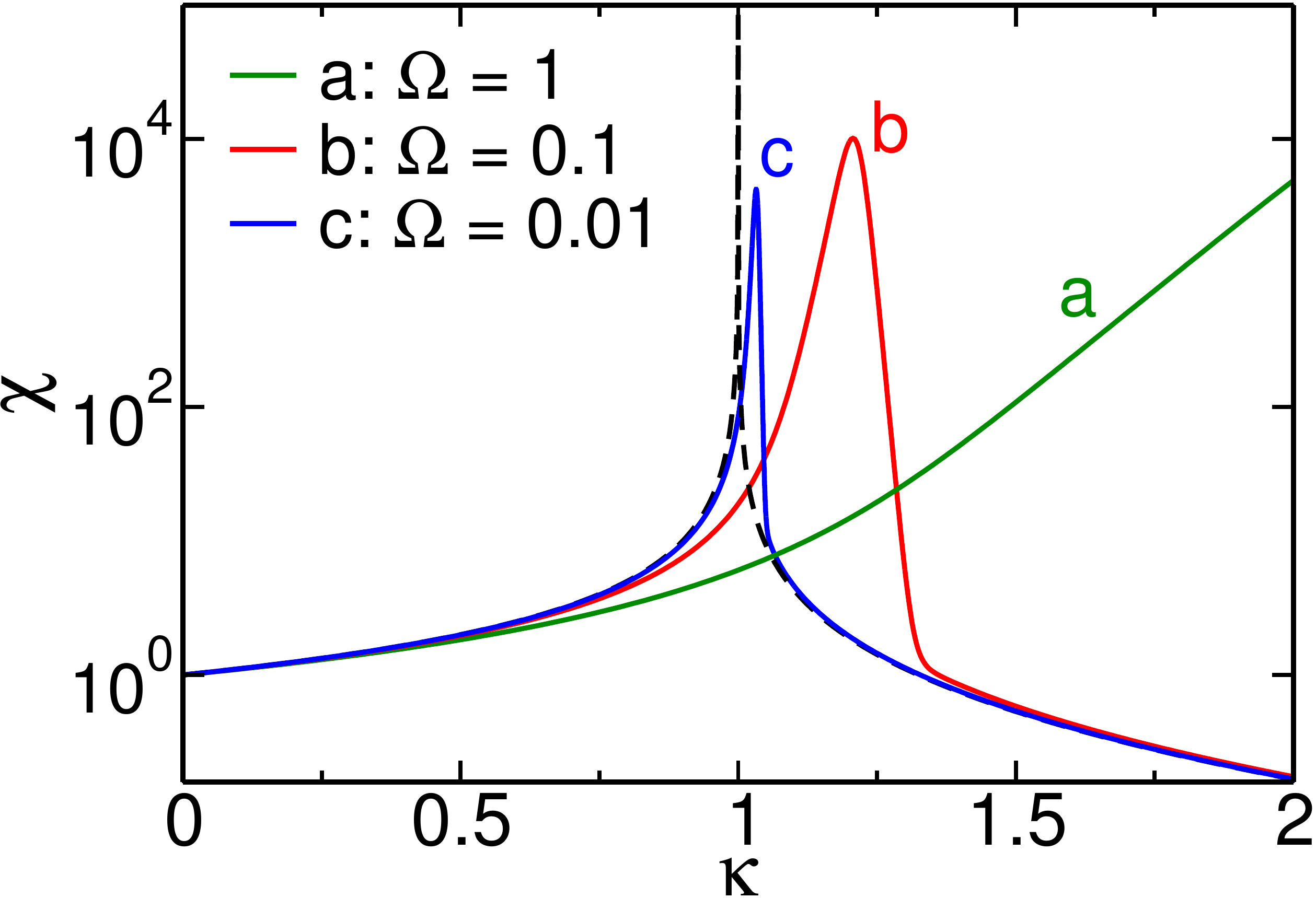} 
\caption{(Color online) QPT in the CO limit. 
Order parameter $\langle J_x \rangle$ (left panel) and
susceptibility $\chi$ (right panel) as a function of $\kappa$,
for various values of $\Omega/\Delta$ and fixed $j=5$.
The dashed curves give the MF result from Eqs.~\eqref{MFJx},~\eqref{MFsus}.}
\label{fig:COQPT}
\end{figure}

Given the QPT in the CS and CO limit,
we expect that the order parameter $\langle J_x \rangle$ and the susceptibility $\chi$
converge to the MF values from Eqs.~\eqref{MFJx},~\eqref{MFsus}
if the classical limits are approached from the quantum regime $j < \infty$, $\Omega/\Delta > 0$, i.e. if one moves in Fig.~\ref{fig:Dicke} from the interior of the square towards one of the two bold axes.

To observe convergence
we must use a small symmetry breaking field $\epsilon J_x$ as in Eq.~\eqref{Heps} to select one of the two possible cases $\langle J_x \rangle \gtrless 0$ in the broken symmetry phase $\kappa>1$.
Without the additional field
convergence cannot be observed because parity symmetry implies 
that $\langle J_x \rangle = 0$ in the quantum regime.
Apart from the calculation of the entanglement entropy in Sec.~\ref{sec:QENT},
we use $\epsilon=10^{-4}$ throughout the paper.

In Figs.~\ref{fig:CSQPT},~\ref{fig:COQPT} we show $\langle J_x \rangle$ and $\chi$ from a numerical calculation of the ground state of the Dicke model
using the Lanczos technique~\cite{So02}.
 Up to $10^{3}$ bosons are kept in the calculations to ensure a negligible error from the truncation of the infinite-dimensional bosonic Hilbert space.
The spin part is not truncated,  and the numerical data are accurate on the level of machine precision.

In the figures we start from the curve for $j=5$, $\Omega/\Delta=1$,
far away from the classical limits.
In Fig.~\ref{fig:CSQPT} we increase $j$ to approach the CS limit,
and in Fig.~\ref{fig:COQPT} we decrease $\Omega/\Delta$ to approach the CO limit.
In both situations 
we observe convergence of $\langle J_x \rangle$, $\chi$
to the MF values from Eqs.~\eqref{MFJx},~\eqref{MFsus}.
Note that in the CO limit convergence takes place while the spin length $j$ remains finite.

\section{Quantum corrections to the mean-field QPT}\label{sec:QFLU}

The previous section showed that MF theory becomes exact in the CS and the CO limit,
which are therefore identical with respect to the critical behavior of the QPT.
The nature of the two limits is however different
and they can be distinguished through the properties of quantum fluctuations around the MF ground state.

The origin of the differences can be understood with a simple energy argument.
In the CO limit, the energy scales for spin fluctuations ($\propto \Delta$)
and oscillators fluctuations ($\propto \Omega$) separate.
Since $\Delta$ is large compared to the coupling constant $\lambda$, which is proportional to $\sqrt{\Omega}$ (cf. Eq.~\eqref{LamTil}), spin fluctuations are suppressed in the CO limit.
This explains partly why the QPT in the CO limit can occur already for finite $j$.
In the CS limit the ratio $\Omega/\Delta$ remains constant and neither spin nor oscillator fluctuations are suppressed.

\subsection{Effective model for the CO limit}\label{sec:EFFCO}

While spin fluctuations are suppressed in the CO limit,
oscillator fluctuations around the classical coherent state $|\alpha\rangle$ remain energetically favorable. 
Their strength can be derived with an effective bosonic model obtained in perturbation theory.

For $\kappa<1$ the MF ground state $|{-j}\rangle \otimes |\mathrm{vac}\rangle$
is the product of the $J_z$--eigenstate $|{-j}\rangle$ to the smallest eigenvalue $-j$
and the bosonic vacuum $|\mathrm{vac}\rangle$.
For $\Delta \gg \Omega$, the low energy sector of the Hilbert space consists of all states
$|{-j}\rangle \otimes |\psi_\mathrm{bos}\rangle$
with a bosonic state $ |\psi_\mathrm{bos}\rangle$.
While the operator $J_z$ remains in the low energy sector,
the operator $J_x$ creates a spin excitation $\propto |{-j+1}\rangle$ of energy $\Delta$.

Standard perturbation theory~\cite{Mes61} gives the effective low energy model
for the bosonic state $|\psi_\mathrm{bos}\rangle$ as
\begin{equation}\label{HBosBelow}
\begin{split}
   H_\mathrm{bos}^< &= \langle{-j} | \Delta J_z + \Omega a^\dagger a + \frac{\lambda^2}{\Delta} \Big[  (a+a^\dagger) J_x \Big]^2  | {-j}\rangle \\
   &= - \Delta j  + \Omega ( a^\dagger a -  \frac{\kappa}{4} (a+a^\dagger)^2 ) \;.
  \end{split}
\end{equation}
Using the results from App.~\ref{app:HOM} for the bosonic part in the second line,
we see that the stability condition in Eq.~\eqref{app:Stability}
is fulfilled only for $\kappa<1$ below the critical coupling.
At the QPT $\kappa=1$ the number of oscillator fluctuations introduced through the term $(a+a^\dagger)^2$ diverges.

For $\kappa>1$ we must consider spin fluctuations above the classical ground state $|\theta\rangle \otimes |\alpha\rangle$,
which are no longer created by $J_x$ since $\theta \ne 0$.
Instead, we rewrite the Hamiltonian with the 
rotated spin operators $\tilde{J}_z = \cos \theta J_z - \sin \theta J_x$,
$\tilde{J}_x = \sin \theta J_z + \cos \theta J_x$,
and find
\begin{multline}\label{HRotate}
  H = \Delta \cos \theta \tilde{J}_z - \lambda \sin \theta \, (a+a^\dagger) \tilde{J}_z + \Omega a^\dagger a \\
  + \Delta \sin \theta \tilde{J}_x + \lambda \cos  \theta \, (a+a^\dagger) \tilde{J}_x \;.
\end{multline}
Here the operator $\tilde{J}_x$ appears with the prefactor $\Delta$,
and we cannot immediately use this expression for perturbation theory for small $\Omega/\Delta$.

To proceed, we shift operators $\tilde{J}_z \mapsto \tilde{J}_z +j $, $a \mapsto a -ÊÊ\alpha$ 
by their expectation values in the classical ground state, taken from Eqs.~\eqref{MFAlpha},~\eqref{MFTheta},
and obtain the Hamiltonian in the form
\begin{multline}
  H = E(\theta) + \Delta (\cos \theta + \kappa \sin^2 \theta) (\tilde{J}_z + j) \\ 
  - \lambda \sin \theta (a + a^\dagger -2 \alpha) (\tilde{J}_z + j) \\
  + \Omega (a^\dagger - \alpha) (a-\alpha)  + \lambda \cos \theta (a+a^\dagger-2 \alpha) \tilde{J}_x \;,
\end{multline}
where the first term is the energy functional $E(\theta)$ from Eq.~\eqref{ErgFunc}.
In this expression, a term $2 \alpha \lambda \cos \theta J_x$ has canceled the problematic term $\Delta \sin \theta \tilde{J}_x$ for the values of $\alpha$, $\theta$ given by  Eqs.~\eqref{MFAlpha},~\eqref{MFTheta}.

Now, $\tilde{J}_x$ in the last line appears with a prefactor that is small compared to $\Delta$ and perturbation theory can be applied.
Note that the spin fluctuation energy, which is given by the prefactor $\Delta (\cos \theta + \kappa \sin^2 \theta) = \Delta \kappa$ of $\tilde{J}_z + j$,
differs from the bare value $\Delta$.
The effective model for $\kappa>1$ is obtained as
\begin{multline}\label{HBosAbove}
 H^>_\mathrm{bos} =  - \frac{j \Delta}{2} \Big(\frac{1}{\kappa} + \kappa \Big) \\ 
 + \Omega \Big[ (a^\dagger -\alpha) (a -\alpha)  -    \frac{1}{4 \kappa^2} (a+a^\dagger-2 \alpha)^2 \Big] \;.
\end{multline} 
For $\kappa\to1$, it coincides with $H_\mathrm{bos}^{<}$ from Eq.~\eqref{HBosBelow}.

The effective low energy models $H^{\gtrless}_\mathrm{bos}$ describe the ground state of the Dicke model in the CO limit including oscillator fluctuations.
From this model we recover the MF expressions for $\alpha$ and $E(\theta)$, and thus the entire QPT in the CO limit. 
In particular the present derivation shows that the argument given in Sec.~\ref{sec:QPTCO} is correct and not  invalidated by oscillator fluctuations.

For a quantitative analysis of the numerical data, we use the oscillator variance
\begin{equation}
 \Delta_q = \langle \hat{q}^2 \rangle -   \langle \hat{q} \rangle^2 
\end{equation}
of the oscillator position $\hat{q}=(a+a^\dagger)$.
Using the results from App.~\ref{app:HOM} the variance 
is obtained from $H^{\gtrless}_\mathrm{bos}$ as
\begin{equation}\label{DeltaX}
 \Delta_q = 
  \begin{cases}
    (1-\kappa)^{-1/2}  \quad & \text{ if } \kappa < 1 \;, \\
    (1-1/\kappa^2)^{-1/2} & \text{ if } \kappa > 1 \;.
  \end{cases}
\end{equation}

\subsection{Spin and oscillator variance}

\begin{figure}
\includegraphics[width=0.48\linewidth]{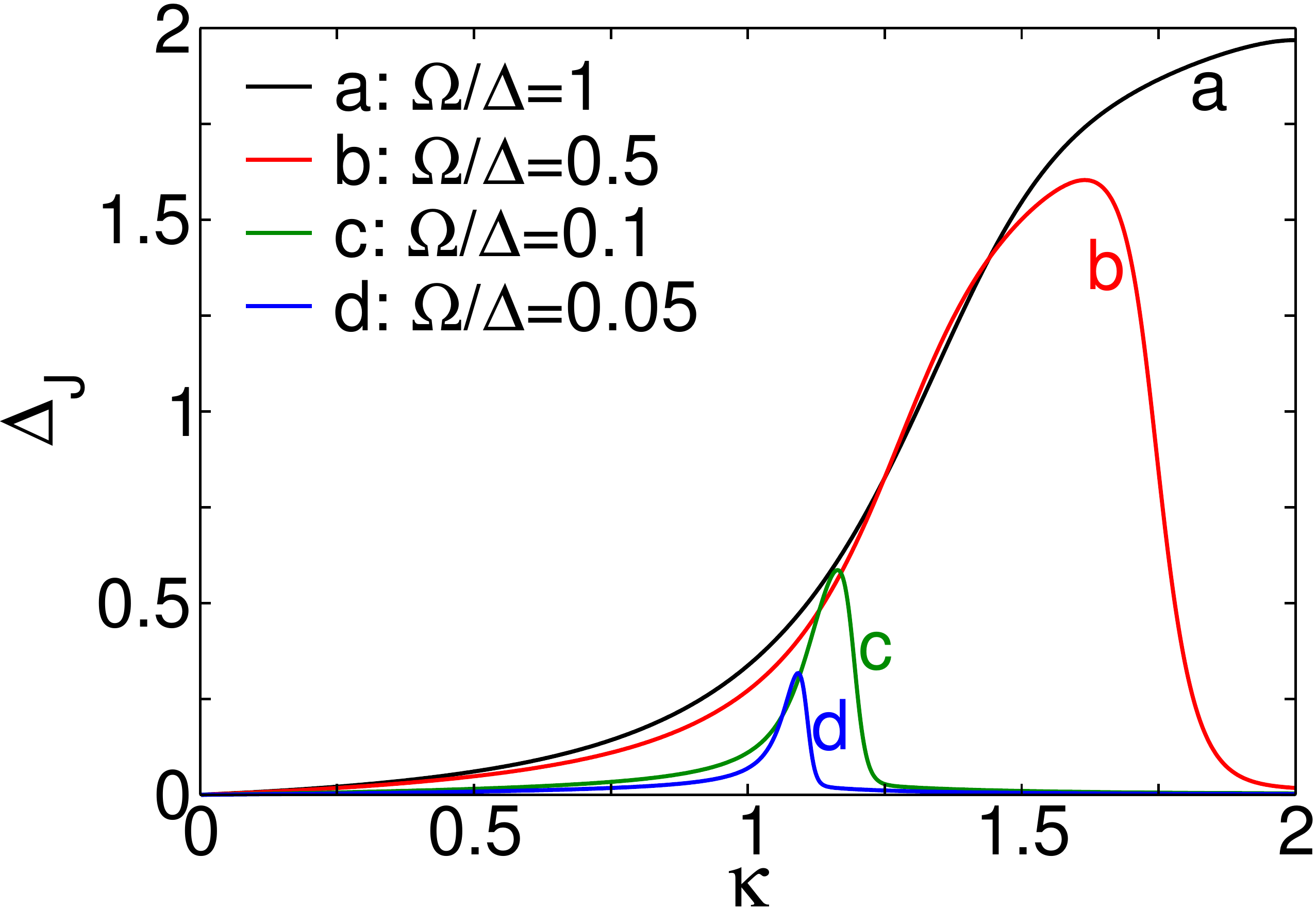} \hfill
\includegraphics[width=0.48\linewidth]{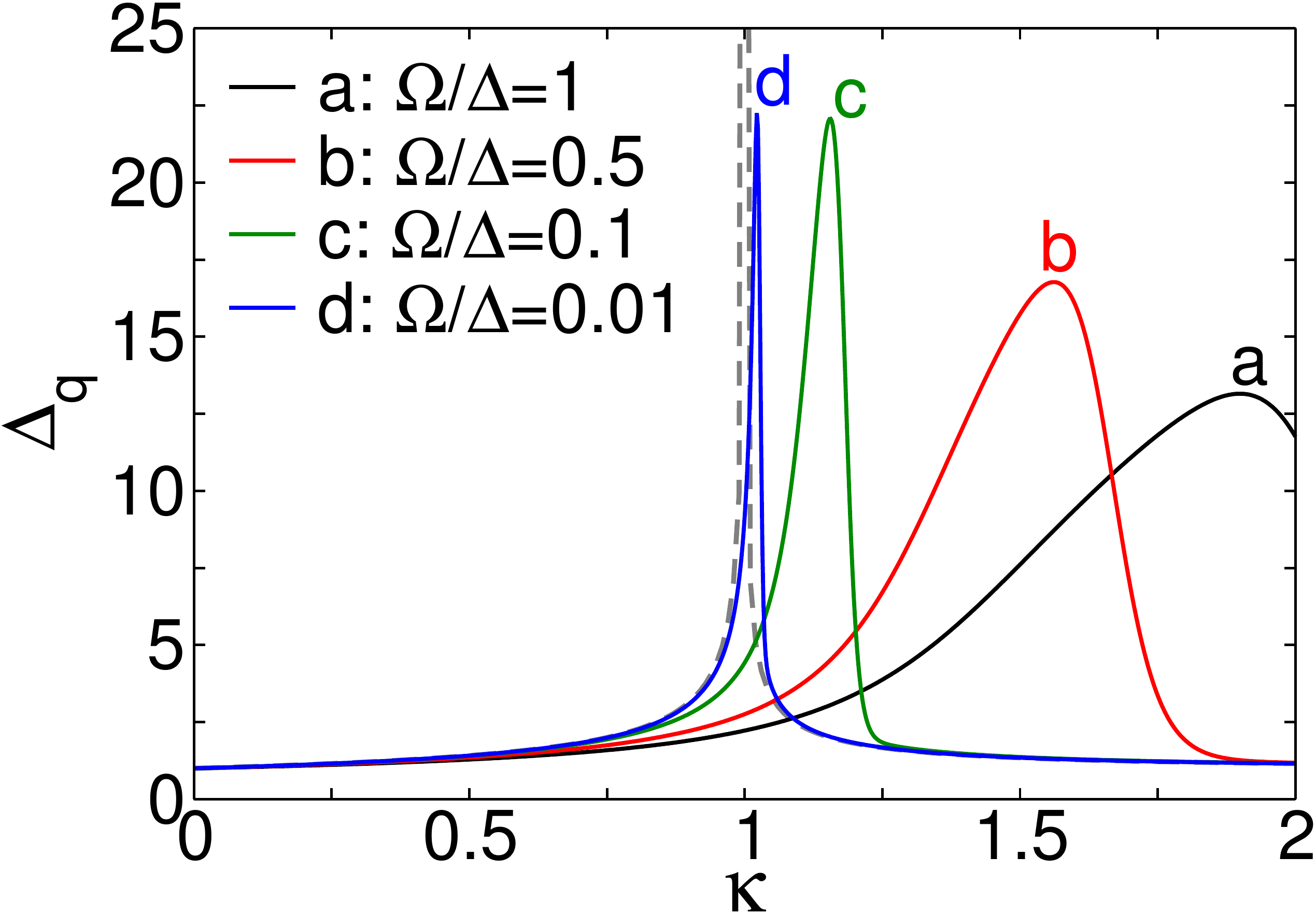} 
\caption{(Color online) Spin variance $\Delta_J$ (left panel) and oscillator variance $\Delta_q$ (right panel)
for decreasing $\Omega/\Delta$ approaching the CO limit, with fixed finite $j=5$.
The dashed grey curve in the right panel gives the analytical for $\Delta_q$ from Eq.~\eqref{DeltaX}.
}
\label{fig:VarCO}
\end{figure}

In Fig.~\ref{fig:VarCO} we compare the oscillator variance $\Delta_q$ 
for small $\Omega/\Delta$ with the spin variance $\Delta_J$.
We use $\Delta_J$, as defined in App.~\ref{app:JVar},
instead of the spin variance in a fixed direction, for example $\Delta_z = \langle J_z^2 \rangle - \langle J_z\rangle^2$,
since it is invariant under rotations.
For fluctuations around the spin coherent state $|\theta=0\rangle$ that points in the $z$-direction it is identical to $\Delta_z$.
Above the QPT,  $\Delta_J$ accounts for the rotation of the spin axis relative to which spin fluctuations occur.
A large $\Delta_J$ is an indication of significant spin fluctuations,
while $\Delta_J=0$ corresponds (in the present examples) to a spin coherent state.
The vanishing of $\Delta_J$ as $\Omega/\Delta \to 0$
(left panel) shows the suppression of spin fluctuations in the CO limit.

For $\Delta_q$ in the right panel we observe the growth of oscillator fluctuations
with decreasing $\Omega/\Delta$. 
Recall that $j=5$ is finite and small in this example,
and the QPT in the CO limit is triggered by a macroscopic displacement of the classical oscillator.
Oscillator fluctuations are a genuine quantum correction, 
which is independent of $j$ and occurs even in the smallest non-trivial case $j=1/2$.
Approaching the CO limit, $\Delta_q$ diverges at the QPT
according to Eq.~\eqref{DeltaX}.
The criticality of quantum fluctuations implies the breakdown of the classical oscillator limit in the vicinity of the QPT.

\section{Critical and non-critical entanglement}\label{sec:QENT}

In addition to the spin and oscillator variance studied in the previous section,
corrections to the MF ground state arise from spin-oscillator entanglement.
It can be measured with the entanglement entropy
\begin{equation}
 S = - \mathrm{Tr} [ \rho \ln \rho] \;,
 \end{equation}
which is calculated with the reduced spin or oscillator density matrix $\rho$
(both choices give the same result according to the Schmidt decomposition)~\cite{NC10,HHHH09}.

A simple argument would suggest a jump of $S$ at $\kappa=1$ from 
$S=0$ for the non-degenerate ground state below the QPT to $S= \ln 2$ for the two-fold degenerate ground state above the QPT.
Note that we assume $\epsilon=0$ here.

Quantum fluctuations can modify this behavior considerably, 
and lead to criticality of entanglement in the CS limit~\cite{LEB04,LEB05}.
In the vicinity of the QPT the entanglement entropy in the CS limit is given as~\cite{LEB05,VDB07} 
\begin{equation}\label{SCS}
 S_\mathrm{CS} =  - \frac{1}{4} \ln |1-\kappa| + \mathrm{const.} \;,
\end{equation}
such that $S_\mathrm{CS} $ diverges at the critical coupling $\kappa=1$ with critical exponent $1/4$.
 
\begin{figure}
\includegraphics[width=0.48\linewidth]{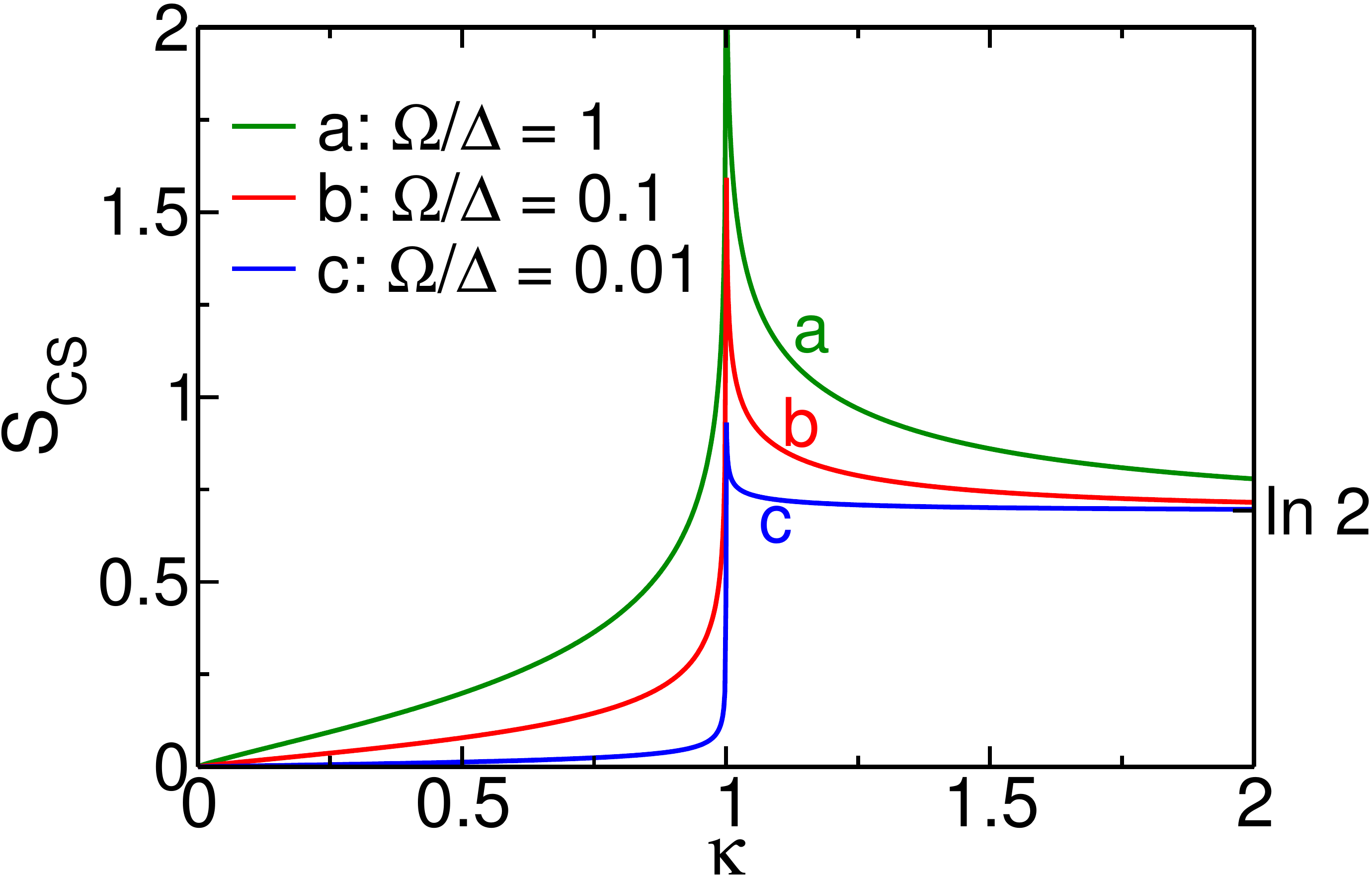} \hfill
\includegraphics[width=0.48\linewidth]{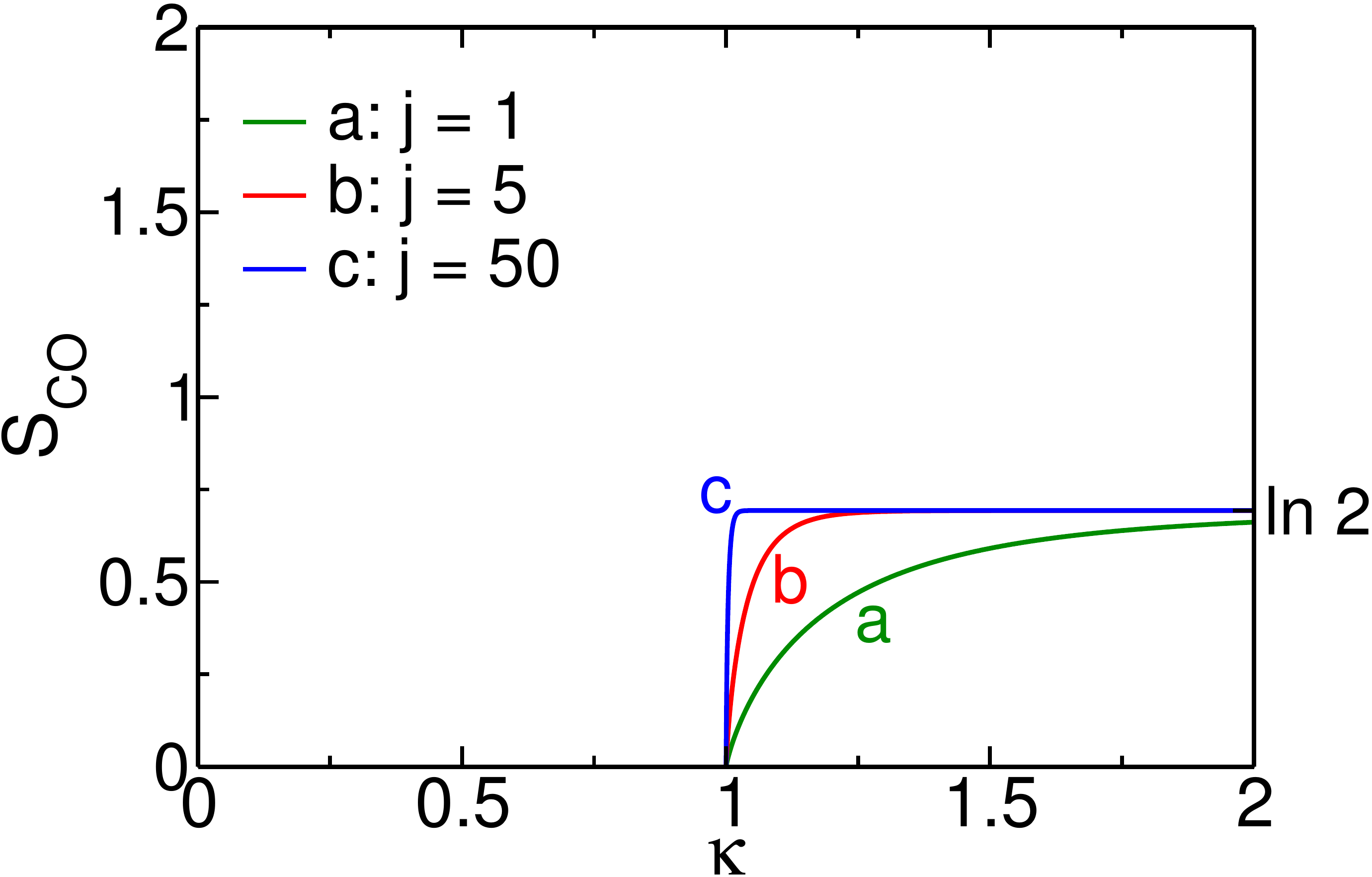} \hfill
\caption{(Color online) Entanglement entropy $S$ in the CS limit (left panel) (cf. Ref.~\cite{LEB04}) and in the CO limit (right panel) according to Eq.~\eqref{SCO}.
}
\label{fig:SClass}
\end{figure}

 We show $S_\mathrm{CS}$ in Fig.~\ref{fig:SClass} (left panel).
 The functional form of $S_\mathrm{CS}=S_\mathrm{CS}(\kappa)$ depends on the ratio $\Omega/\Delta$.
 For $\Omega/\Delta \ll 1$ or  $\Omega/\Delta \gg 1$ 
 quantum spin or oscillator fluctuations are energetically less favorable than for $\Omega/\Delta=1$, and the value of $S$ decreases away from the QPT.
The criticality of $S_\mathrm{CS}$ at the QPT and the critical exponent are however independent of $\Omega/\Delta$.

\subsection{Entanglement in the classical oscillator limit}

Almost trivially, the entanglement entropy $S$ cannot diverge 
for finite $j$ since it is bounded by $S \le \ln (2 j +1)$. 
The suppression of spin fluctuations in the CO limit 
results in the much stricter condition $S \le \ln 2$, independently of $j$.
Large entanglement requires a sizeable amount of both oscillator and spin fluctuations.

For $\kappa<1$ the ground state is a product state and $S_\mathrm{CO} [\kappa < 1] =0$.
This follows immediately from the fact used in the perturbative calculation in Sec.~\ref{sec:EFFCO}
that the spin part of the ground state is the single $J_z$-eigenstate $|-j\rangle$.
In contrast to the CS limit, the absence of spin fluctuations prevents the growth of $S_\mathrm{CO}$ 
with increasing $\kappa$.

For $\kappa>1$ the symmetrized ground state wave function is
given by
\begin{equation}
|\psi\rangle = \dfrac{1}{\sqrt{2}} \Big( |\theta \rangle \otimes |\alpha\rangle \pm  |{-\theta }\rangle \otimes |{-\alpha}\rangle \Big) \;.
\end{equation}
Note that the we assume $\epsilon=0$ here.

The two states in the bracket are orthogonal since $\langle \alpha| -\alpha\rangle=0$ in the CO limit ($\alpha$ diverges according to Eq.~\eqref{MFAlpha}),
but the two spin coherent states $|{\pm \theta}\rangle$ are not orthogonal
such that $S$ remains strictly smaller than $\ln 2$.

The reduced spin density matrix is
\begin{equation}
\rho_s = \frac{1}{2} \Big(|\theta\rangle\langle\theta| + |{-\theta}\rangle\langle-\theta|  \Big) \;,
\end{equation}
with eigenvalues 
$\mu_{\pm} = 1 \pm \langle\theta|{-\theta}\rangle= 1 \pm \cos^{2j} \theta$.
Note that this expression for $\rho_s$ is valid also for the non-orthogonal states appearing here.

The entanglement entropy obtained from $\rho_s$ as
$S_\mathrm{CO} = - \mu_- \ln \mu_-  \, -  \mu_+ \ln \mu_+$
is 
\begin{multline}\label{SCO}
S_\mathrm{CO} [\kappa>1] = \ln 2 \, - \frac{1}{2} (1-\kappa^{-2j}) \ln (1-\kappa^{-2j}) \, \\ - \frac{1}{2} (1+\kappa^{-2j}) \ln (1+\kappa^{-2j}) \; ,
\end{multline}
where we inserted the angle $\cos \theta=1/\kappa$ according to Eq.~\eqref{MFTheta}.

We show $S_\mathrm{CO}$ in Fig.~\ref{fig:SClass} (right panel),
where it can be compared to $S_\mathrm{CS}$.
In the CO limit the entropy remains zero for $\kappa<1$ and increases monotonically from $0$ to $\ln 2$ for $\kappa>1$.
Note that $S_\mathrm{CO} \le \ln 2$ for all $j$ and $\kappa$.

For finite $j$, it remains $S_\mathrm{CO}< \ln 2$ even above the QPT
because of the finite overlap $\langle \theta | {- \theta} \rangle$.
For $j \to \infty$, still strictly in the CO limit $\Omega=0$,
the overlap of the two spin coherent states vanishes and the curves approach a step function with a jump at $\kappa=1$.

\subsection{Quantum regime}

\begin{figure}
\includegraphics[width=0.48\linewidth]{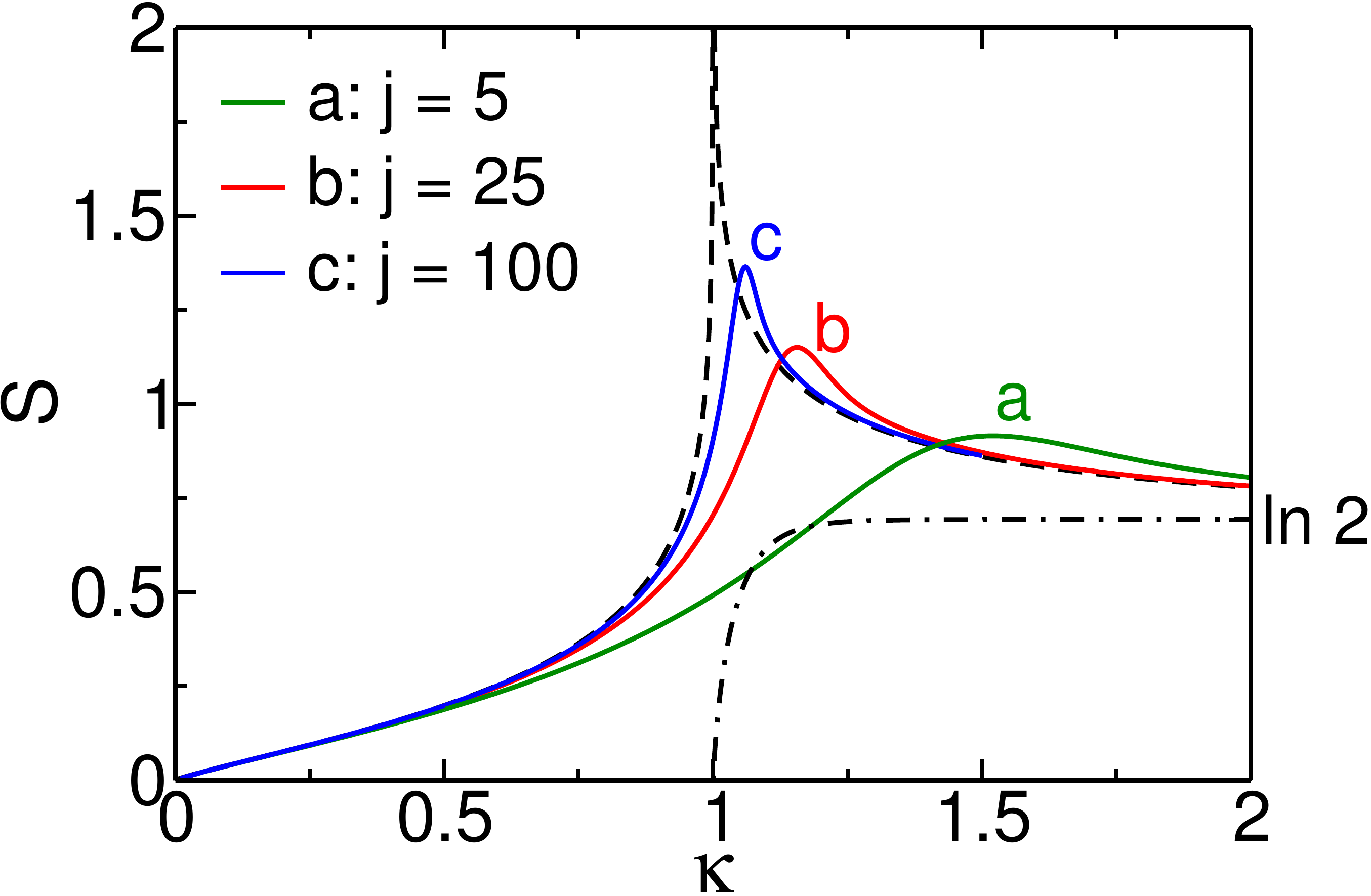} \hfill
\includegraphics[width=0.48\linewidth]{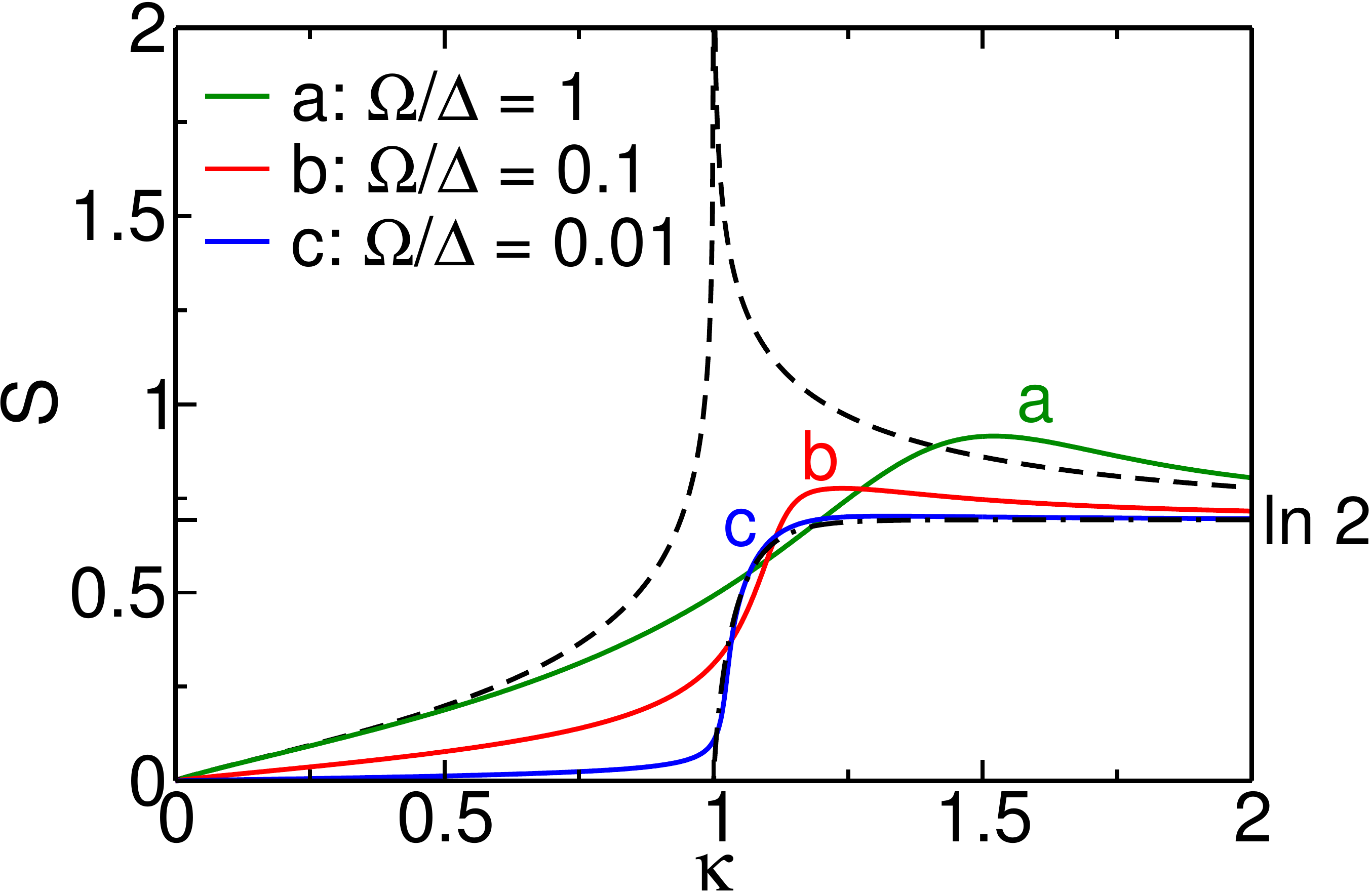} \\[0.5ex]
\caption{(Color online) Convergence of entanglement entropy $S$ towards the classical limits.
Left panel: Approaching the CS limit with increasing $j$ for fixed $\Omega/\Delta=1$.
Right panel: Approaching the CO limit with decreasing $\Omega/\Delta$ for fixed $j=5$.
Both panel start from the curve for $j=5$, $\Omega/\Delta=1$.
The dashed (dot-dashed) curve gives the analytical result in the CS (CO) limit according to Fig.~\ref{fig:SClass} and Eqs.~\eqref{SCS},~\eqref{SCO}. 
}
\label{fig:SQuant1}
\end{figure}

We show in Fig.~\ref{fig:SQuant1} the convergence of the entanglement entropy $S$
to the analytical results for the two classical limits.
As in Figs.~\ref{fig:CSQPT},~\ref{fig:COQPT}
we start from the curve for $j=5$, $\Omega/\Delta=1$.
In the CS limit (left panel) we see how the divergence of $S$ at $\kappa=1$ develops
as $j$ is increased.
For the CO limit (right panel) we see that $S$ remains small 
and converges to a continuous function bounded by $\ln 2$.
Since $j$ is finite, the limiting curve is continuous without the jump at $\kappa=1$ that evolves only for $j \to \infty$.

It should be noted that the CO limit with $j \gg 1$ and small entanglement differs from the regime $\Omega \ll \Delta$ in the CS limit where the entanglement entropy still diverges at the QPT~\cite{LZPP06,LPP06}.
Both situations are located close to the origin in Fig.~\ref{fig:Dicke},
but the first (second) situation lies closer to the abscissa (ordinate) than to the second axis.

\section{Fast oscillator limit and the Lipkin-Meshkov-Glick model}
\label{sec:FO}

Opposed to the CO limit is the fast oscillator (FO) limit $\Omega/\Delta \rightarrow \infty$. 
In this limit oscillator fluctuations are suppressed,
while the spin fluctuations are described by an effective model that can be derived in perturbation theory analogously to Sec.~\ref{sec:EFFCO}.
This results in the Lipkin-Meshkov-Glick (LMG) model~\cite{LMG65} known from nuclear physics.
For the special case $j=1/2$, where the Dicke model reduces to the Rabi model,
the FO limit can also be performed with a different scaling of the coupling constant.

\subsection{Derivation of the LMG model in the FO limit}

We can derive the LMG model as the effective low energy model for the spin part of the wave function
in analogy to Sec.~\ref{sec:EFFCO}.
The derivation is in fact easier than in the CO limit, 
since fluctuations around the bosonic coherent state $|\alpha\rangle$ are described by translated bosonic operators $a^{\dagger}-\alpha$ instead of rotated (spin) operators.
The effective spin model, which is valid for all $\kappa$, is obtained as the LMG model
\begin{equation}\label{LMG}
 H_\mathrm{LMG} = \Delta \Big( J_z  - \frac{ \kappa}{2j} J_x^2 \Big) \;.
\end{equation}

\begin{figure}
\includegraphics[width=0.48\linewidth]{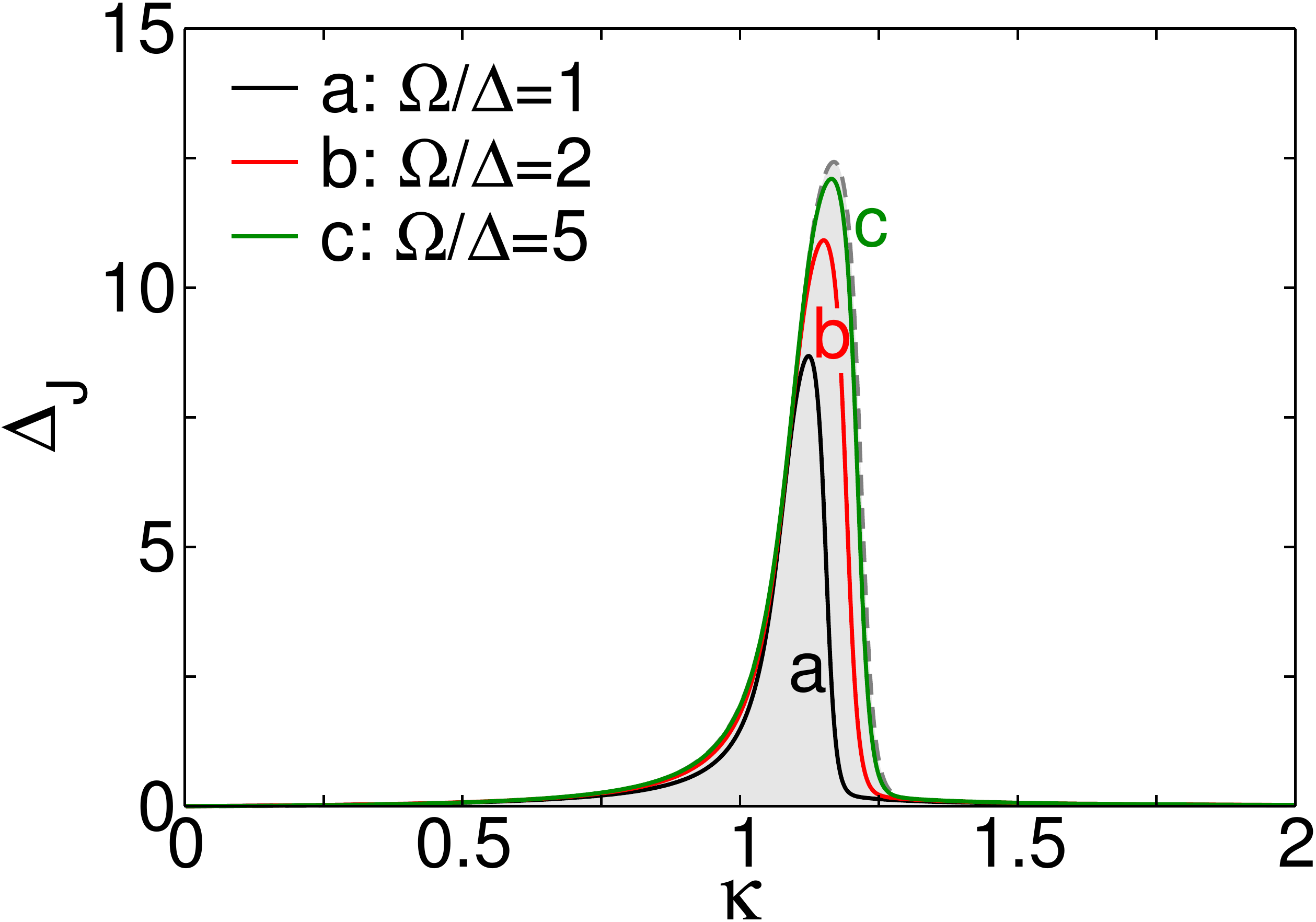} \hfill
\includegraphics[width=0.48\linewidth]{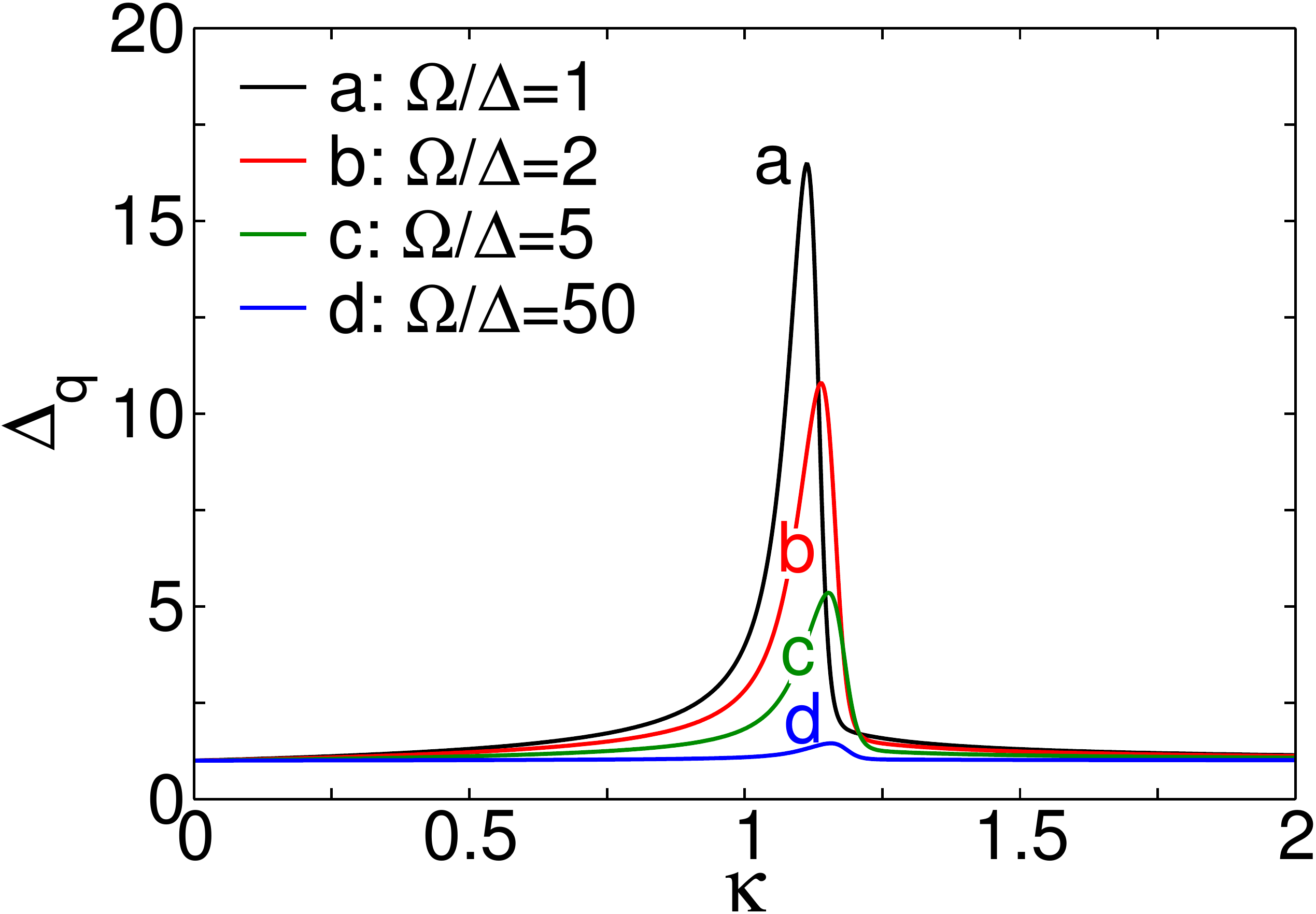} 
\caption{(Color online) Spin variance $\Delta_J$ (left panel) and oscillator variance $\Delta_q$ (right panel)
for increasing $\Omega/\Delta$ approaching the FO limit, with fixed large $j=50$.
The grey filled background curve in the left panel gives $\Delta_J$ for the LMG model.
This figure complements Fig.~\ref{fig:VarCO} for the CO limit.
}
\label{fig:VarFO}
\end{figure}

In Fig.~\ref{fig:VarFO}
we show the spin variance $\Delta_J$ (left panel) and the oscillator variance $\Delta_q$ (right panel) for large $\Omega/\Delta$.
In reversal of the behavior in the CO limit,
we see the suppression of oscillator fluctuations.
Spin fluctuations remain finite and converge for $\Omega/\Delta \to \infty$ to the result from the LMG model.

In contrast to the CO limit no QPT and, therefore, no divergence of spin fluctuations
occurs in the FO limit for finite $j$,
 simply because the initial argument against symmetry breaking given in the introduction applies to the LMG model.
The QPT is recovered if additionally the $j\to\infty$-limit is performed in the LMG model.
Then, the ground state of $ H_\mathrm{LMG}$ is a spin coherent state $|\theta\rangle$, and we recover the energy functional $E(\theta)$ from Eq.~\eqref{ErgFunc}.
Consequently, we also recover the QPT.
In this sense, the FO and CS limit commute.

To calculate the spin fluctuations for $j \to \infty$, we use the Holstein-Primakoff (HP) transformation~\cite{HP40}
\begin{equation}
 J_z = b^\dagger b - j \;, \quad J_x = \sqrt{\frac{j}{2}} ( b^\dagger +b ) + O(j^{-1/2})
\end{equation}
of spin operators to bosonic operators $b^{(\dagger)}$.
For $\kappa<1$, when $\theta=0$, the HP transformation can be applied directly to $H_\mathrm{LMG}$ and results in the bosonic model
\begin{equation}
 H^<_{\infty} = \Delta \left(  b^\dagger b - \frac{\kappa}{4 } (b+b^\dagger)^2  - j   \right)  \;.
\end{equation}

For $\kappa>1$ it is $\theta \ne 0$ and spin operators must be rotated prior to the HP transformation, similarly to Eq.~\eqref{HRotate} in Sec.~\ref{sec:EFFCO}.
We now obtain the bosonic model
\begin{equation}
 H^>_{\infty} = \Delta \kappa \left(  b^\dagger b - \frac{1}{4 \kappa^2} (b+b^\dagger)^2 \right)  - \frac{j \Delta}{2} ( \kappa + 1/\kappa)  \;.
\end{equation}
Comparison of these models to the effective bosonic models $H^\gtrless_\mathrm{bos}$ from Eqs.~\eqref{HBosBelow},~\eqref{HBosAbove} reveals the duality of the CO limit and the combined FO/CS limit, in the sense that the role of spin and oscillator fluctuations are reversed.

\begin{figure}
\centering
\includegraphics[width=0.48\linewidth]{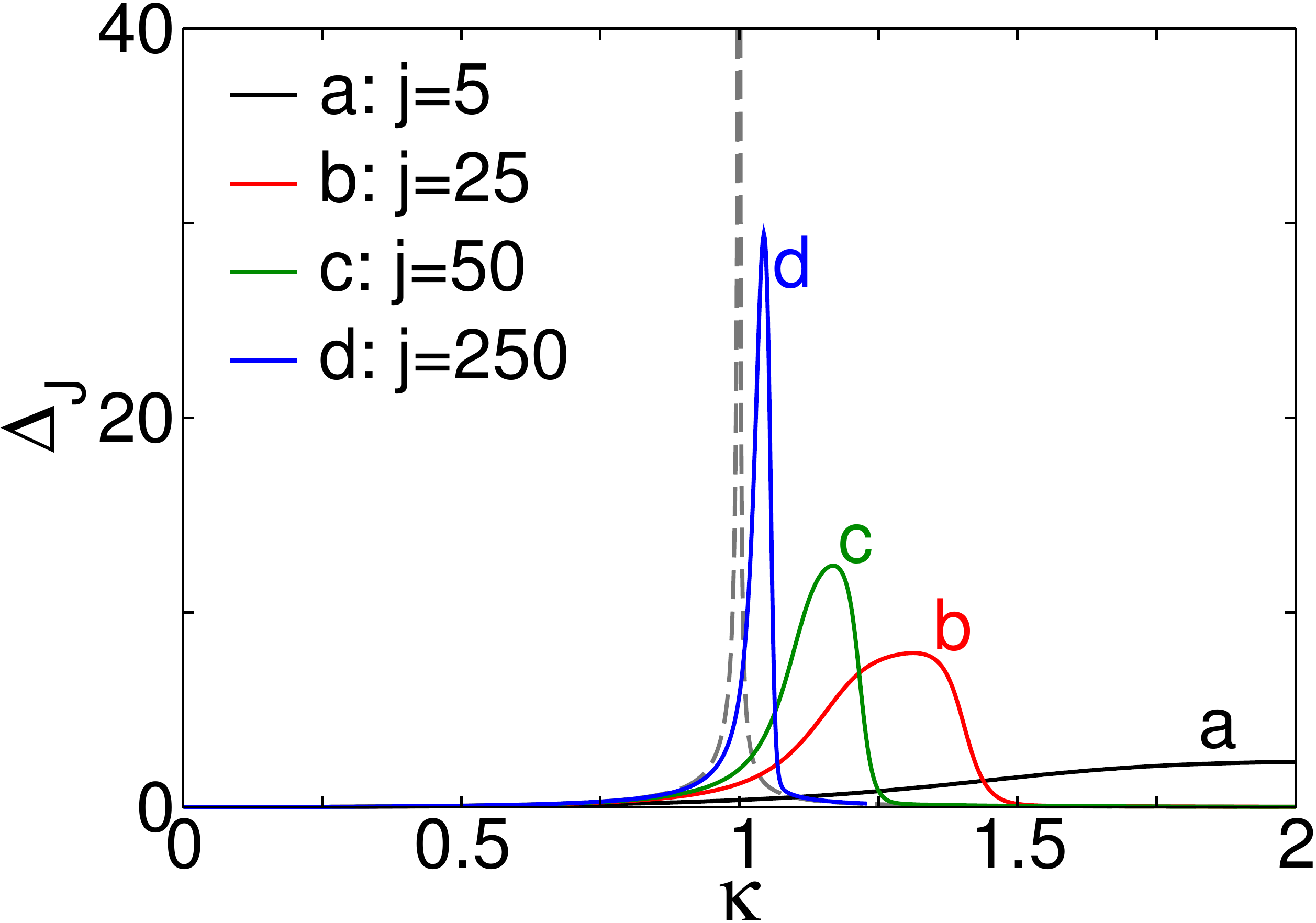} 
\caption{(Color online) Spin variance $\Delta_J$ 
for increasing $j$ approaching the QPT, with fixed large $\Omega/\Delta=20$.
The grey dashed curve gives the analytical result for $\Delta_J$ from Eq.~\eqref{DeltaJFO}.
}
\label{fig:VarFO2}
\end{figure}

Using the results from App.~\ref{app:HOM} for $H^\gtrless_\infty$, we find the spin variance $\Delta_J$ as (still for $j \to \infty$)
\begin{equation}\label{DeltaJFO}
 \Delta_J^\infty =
  \begin{cases}
    \dfrac{\kappa^2}{ 8 (1-\kappa)}  \quad & \text{ if } \kappa < 1 \;, \\[4ex]
   \dfrac{1}{8 \kappa^2 (\kappa^2-1)}  & \text{ if } \kappa > 1 \;.
  \end{cases} 
\end{equation}
The spin variance for large $j$ is shown in Fig.~\ref{fig:VarFO2}.
We observe convergence to the analytical result for $j \to \infty$.
Again, the QPT is accompanied by a divergence of fluctuations and a breakdown of the 
corresponding classical spin limit.

\subsection{FO limit for the Rabi case $j=1/2$}

The FO limit can also be treated by a unitary transformation $U H U^\dagger$ of the Dicke Hamiltonian, where  $U$ is given as
\begin{equation}\label{UTrafo}
U=\exp \Big[ - \xi (\ad-a) J_x \Big] \;, \quad \xi = \frac{\lambda}{\Omega}  \;.
\end{equation}
The transformation $U$ displaces oscillator states by a shift $m_x \xi$ that depends on the $J_x$-eigenvalues $m_x$. 
The equivalent transformation in polaron physics is known as the Lang-Firsov transformation~\cite{LF62r}.

With the above choice for $\xi$, the interaction term $(a^\dagger+a) J_x$ is eliminated through the transformation, and the transformed Hamiltonian reads
\begin{equation}\begin{split}
  U H U^\dagger =& \Delta 
  \cosh [ \xi (a^\dagger-a) ] J_z + \ii \Delta \sinh [ \xi (a^\dagger-a) ] J_y \\
   &+ \Omega a^\dagger a - \frac{\lambda^2}{\Omega} J_x^2 \;.
  \end{split}
\end{equation}

So far, the transformation is an exact reformulation of the problem.
We can now note that for $\Omega \to \infty$ the presence of the term $\Omega a^\dagger a$ 
implies that the ground state of $U H U^\dagger$ contains no bosonic excitations.
The transformed ground state wave function has the form $|\psi_\mathrm{spin}\rangle \otimes |\mathrm{vac}\rangle$.
In the vacuum $|\mathrm{vac}\rangle$ the bosonic operators from $U J_z U^\dagger$ have expectation values
\begin{equation}
\begin{split}
 \langle \mathrm{vac}|  \cosh [ \xi (a^\dagger-a) ]  |Ê\mathrm{vac}\rangle &= e^{-\xi^2/2} \;, \\ 
 \langle \mathrm{vac}|  \sinh [ \xi (a^\dagger-a) ]  |Ê\mathrm{vac}\rangle &=  0  \;,
\end{split}
\end{equation}
and the transformed Hamiltonian $U H U^\dagger$ reduces to 
\begin{equation}
   H_\mathrm{FO} = \Delta e^{-(\lambda^2/\Omega^2)/2} J_z - \frac{\lambda^2}{\Omega} J_x^2 \;.
\end{equation}

We can perform the FO limit in this model in two relevant ways.
If we insert the coupling constant $\kappa$ from Eq.~\eqref{LamTil} as done before,
we obtain the LMG model from Eq.~\eqref{LMG}.
Alternatively, we can keep the parameter $\xi=\lambda/\Omega$ of the transformation constant.
Then, the prefactor of $J_x^2$ will diverge for $\Omega \to \infty$ as we push the system into the strong coupling limit above the QPT.
This limit is not interesting for general $j$.

In the special Rabi case $j=1/2$, however, it is $J_x^2 = 1/4$.
The first kind of FO limit is trivial for this model, since it results in the Hamiltonian $\Delta J_z - \kappa/4$ of a free spin.
Instead, we can perform the second FO limit because the divergent prefactor of $J_x^2$ now results only in a divergent shift of the ground state energy that can be dropped from the Hamiltonian.

We then obtain the simple model
\begin{equation}\label{RabiRen}
 H_\mathrm{ren} = \tilde{\Delta} J_z 
\end{equation}
of a spin $J_z$ with renormalized frequency $\tilde{\Delta} = e^{-\xi^2/2} \Delta$.
It results in the susceptibility
\begin{equation}\label{ChiRabi}
\chi_\mathrm{ren} = \tilde{\Delta}^{-1} = \dfrac{e^{\xi^2/2}}{\Delta} \;,
\end{equation}
which grows monotonically with the effective coupling strength $\xi = \lambda/\Omega$.
The frequency renormalization is the sole effect of coupling to the fast oscillator.
The exponential prefactor is an example of a Franck-Condon factor,
known from the theory of vibronic transitions or polaron physics.

While the transformed model $H_\mathrm{ren}$ is also not very interesting by itself,
with a simple ground state $|{-\nicefrac{1}{2}}\rangle  \otimes |\mathrm{vac}\rangle$,
it allows us to obtain the actual ground state of the Rabi model through the transformation $U$. It is given by
\begin{equation}
\begin{split}
 |\psi\rangle &= U \Big[  |{-\nicefrac{1}{2}}\rangle  \otimes |\mathrm{vac}\rangle \Big] \\
 &= \frac{1}{\sqrt{2}} ( |{\rightarrow}\rangle \otimes | \xi \rangle - |{\leftarrow}\rangle \otimes |{-\xi}\rangle) \;,
 \end{split}
\end{equation} 
where we denote the $j=\pm 1/2$ eigenstates of $J_x$ by $|{\rightarrow}\rangle$, $|{\leftarrow}\rangle$.
The entanglement entropy in this state is given by
\begin{multline}\label{SRabi}
S_\mathrm{ren}  = 
\ln 2 \, - \frac{1}{2} (1- e^{-2 \xi^2}) \ln (1-e^{-2 \xi^2}) \, \\ - \frac{1}{2} (1+e^{-2 \xi^2}) \ln (1+e^{-2 \xi^2}) \; ,
\end{multline}
(cf. the calculation for Eq.~\eqref{SCO}).
Similar to the susceptibility, the entropy $S$  increases monotonically with $\xi = \lambda/\Omega$ from $0$ to $\ln 2$.

\begin{figure}
\includegraphics[width=0.48\linewidth]{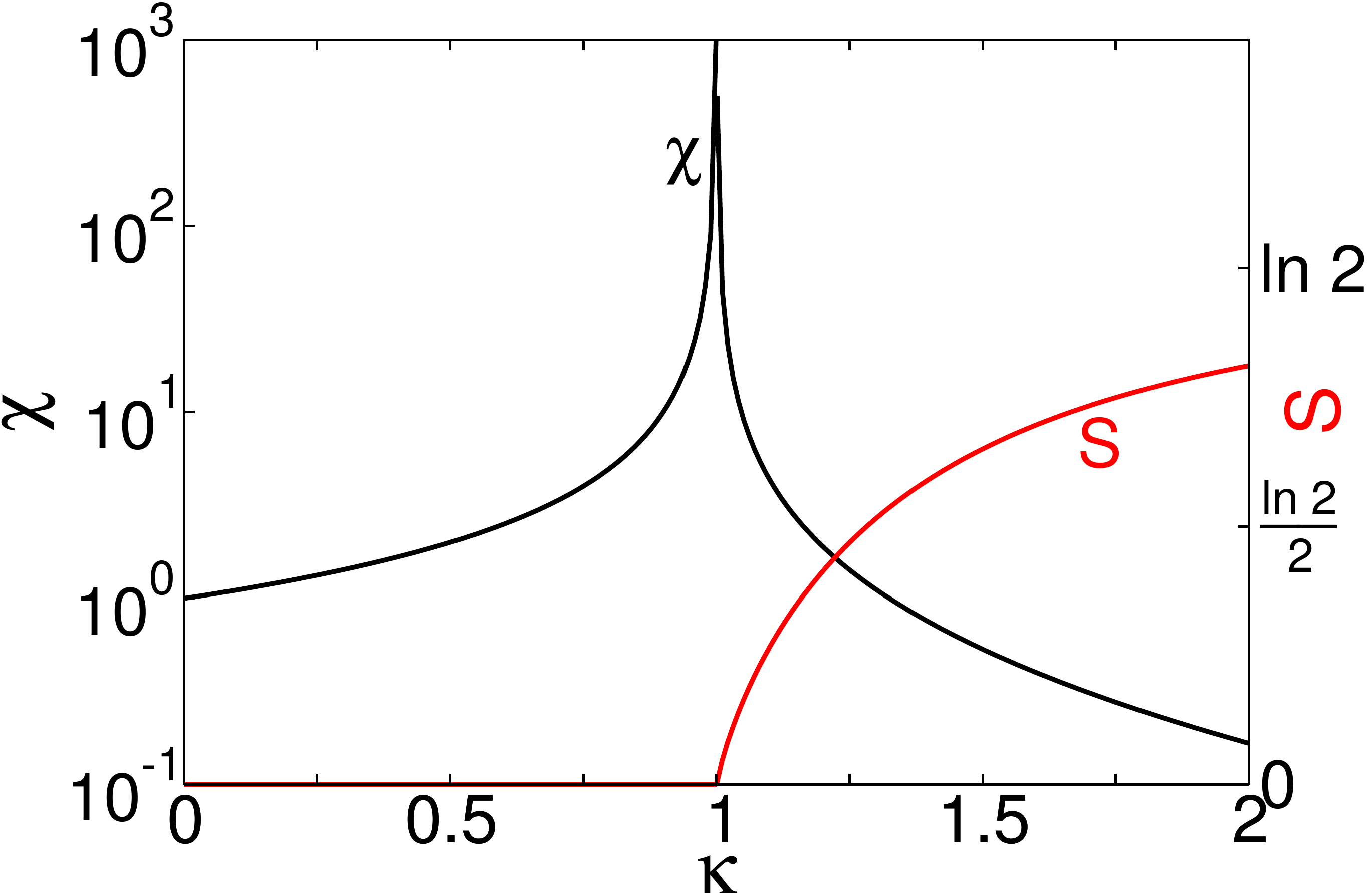} \hfill
\includegraphics[width=0.48\linewidth]{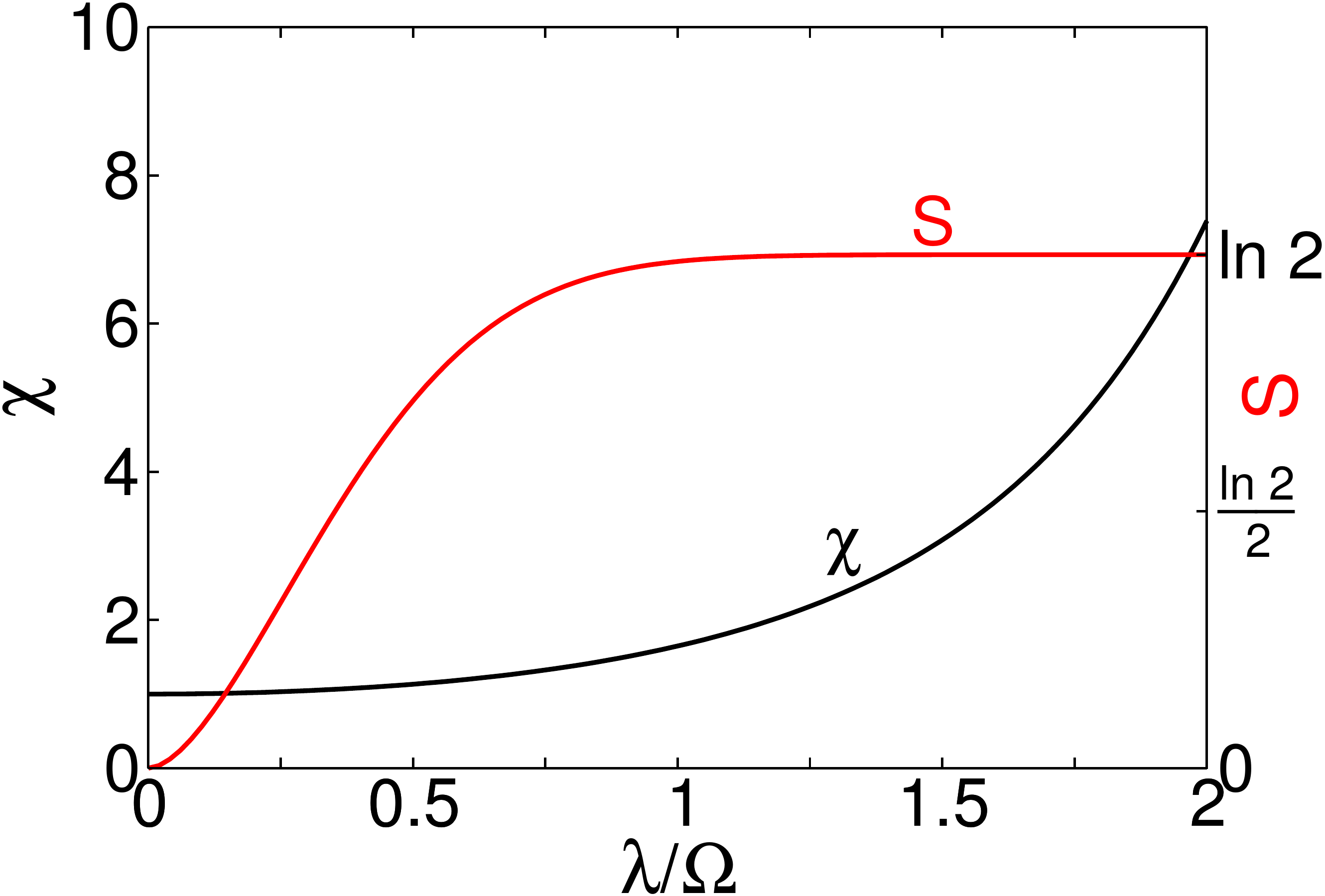} \hfill
\caption{(Color online) Susceptibility $\chi$ and entanglement entropy $S$
for the Rabi case $j = \nicefrac{1}{2}$ in the CO limit (left panel, Eqs.~\eqref{MFsus},~\eqref{SCO}) and the FO limit (right panel, Eqs.~\eqref{ChiRabi},~\eqref{SRabi}).}
\label{fig:RenRabi}
\end{figure}

In Fig.~\ref{fig:RenRabi} we contrast this behavior with the behavior in the CO limit (also for $j=1/2$).
We see that the entanglement entropy $S$ is close to its maximal value $\ln 2$ already for small $\chi$ close to one. In this sense, the FO limit of the Rabi model is characterized by significant entanglement.

Note that the relevant coupling constant scales as $\lambda/\Omega$ in the present FO limit but as $\lambda^2/\Omega$ in the CO limit.
Previously, for the Dicke model, we had to choose the same coupling constant $\kappa \propto \lambda^2/\Omega$ for both limits, which leads to a duality of the FO and CO limit.
For the Rabi model, we have defined a FO limit that is structurally different
from the CO limit: Instead of a QPT it features renormalization of the spin frequency. 

The frequency renormalization is peculiar for the Rabi model.
It is reminiscent of lattice polarons,  where the distinction between self-trapped adiabatic polarons (corresponding to the CO limit) and anti-adiabatic polarons (in the FO limit) involves characteristically different signatures in, e.g., the optical conductivity~\cite{AFT10}.
In fact, the Rabi model is equivalent to the Holstein polaron model restricted to two lattice sites~\cite{RT92}.

\section{Summary}

Two different classical limits can be defined for the Dicke model.
The QPT in the CS spin limit $j \to \infty$ has attracted much attention,
and the criticality of spin-oscillator entanglement is understood as its characteristic signature.

As pointed out here the QPT is also realized in the second classical limit, the CO limit $\Omega \to 0$. 
It should be noted that the QPT in the CO limit is not simply a special case of the CS limit:
It occurs already at finite spin length $j < \infty$, even at $j=1/2$.

A simple MF argument shows the equivalence of the QPT in the two limits if only the critical behavior is considered.
Differences occur for quantum fluctuations around the MF ground state.
In the CO limit, the suppression of spin fluctuations prevents significant entanglement,
but oscillator fluctuations are important and diverge at the QPT.

The emergence of a QPT in the CO limit is a general feature,
which can occur for any finite quantum system coupled to a harmonic oscillator.
It does not require the equivalent of an $j \to \infty$--limit.
In every situation the QPT is accompanied by diverging oscillator fluctuations,
while fluctuations of the finite system are suppressed.

Also the FO limit can be performed for general systems,
although a different formulation of the limit should be chosen for two-level systems (the Rabi case $j=1/2$) and systems with multiple energy levels.
The duality of the CO and FO limit  is a special feature of the Dicke model,
where the spin can be mapped onto a bosonic system for $j\to\infty$.

The basic physical mechanisms realized in the CO and FO limit are typical for any finite quantum system coupled to harmonic oscillators.  Oscillator fluctuations or entanglement with the oscillator for two-level systems in the FO limit can be expected to be of general importance in many situations. The Dicke model is an example where their properties can be studied in detail.
One particular feature is the occurrence of one QPT in two classical limits,
which are distinguished by entirely different quantum corrections to the MF ground state.

\begin{acknowledgments}
This work was supported by 
DFG through  AL1317/1-2 and SFB 652.
\end{acknowledgments}

\appendix

\section{Squeezed harmonic oscillator}\label{app:HOM}

The Hamiltonian of a squeezed harmonic oscillator has the form
\begin{equation}
  H = a^\dagger a + \beta (a+a^\dagger)^2
\end{equation}
with $\beta \in \mathbb{R}$.

The Hamiltonian can be diagonalized with a unitary transformation
\begin{equation}
 U = \exp \Big[ \frac{\sigma}{2} (a^\dagger -a)  \Big] \;,
\end{equation}
for which
\begin{equation}
 U a U^\dagger = \cosh \sigma \, a - \sinh \sigma \, a^\dagger \;.
\end{equation}
With the choice
\begin{equation}
 \tanh 2 \sigma = \frac{2 \beta}{1 + 2 \beta} 
\end{equation}
the transformed Hamiltonian 
\begin{equation}
 \tilde{H} = U  H U^\dagger  = \sqrt{1 + 4 \beta} \, a^\dagger a \; + E_0 
\end{equation}
acquires the form of a standard harmonic oscillator Hamiltonian,
with ground state energy
\begin{equation}
 E_0 =   \frac{1}{2} \sqrt{1 + 4 \beta} - \frac{1}{2}  - \frac{\alpha^2}{1 + 4 \beta} \;.
\end{equation}
We note the stability condition
\begin{equation}\label{app:Stability}
   \beta > - \frac{1}{4} \;.
\end{equation}
For smaller $\beta$, the original Hamiltonian is not bounded from below.

Since the ground state of $\tilde{H}$ is the bosonic vacuum,
expectation values $\langle \dots \rangle$ in the ground state of $H$ can be evaluated through transformation with $U$.
Especially for the oscillator variances we find
\begin{equation}
 \langle (a+a^\dagger)^2 \rangle - \langle a + a^\dagger \rangle^2 = \frac{1}{\sqrt{1 + 4 \beta}}
\end{equation}
and 
\begin{equation}
 \langle (a^\dagger a)^2 \rangle - \langle a^\dagger a\rangle^2 = \dfrac{2 \beta^2}{1+4\beta} \;.
\end{equation}

\section{Some properties of coherent states} \label{app:Coh}
We summarize the essential properties of oscillator and spin coherent states, see also Ref.~\cite{ZFG90}. 

\subsection{Oscillator coherent states}

Coherent states of the oscillator can be defined as the eigenstates of the destruction operator.
\begin{equation}\label{app:alpha}
a |\alpha\rangle = \alpha |\alpha\rangle
\end{equation}

The coherent state can be written as
\begin{equation}
|\alpha\rangle = e^{\alpha \ad - \alpha^* a} |0\rangle
\end{equation}
Expectation values are given by
\begin{align}
\langle \alpha | \ad a | \alpha \rangle &= |\alpha|^2 \notag \\
\langle \alpha | \ad + a | \alpha \rangle &= 2 \Re \alpha
\end{align}

\subsection{Spin coherent states}

A (real) spin coherent state is defined as
\begin{equation}\label{app:theta}
|\theta\rangle = e^{\ii \theta J_y} |j,-j\rangle \;.
\end{equation}
It is the eigenstate of the operator $\cos \theta J_z  - \sin \theta  J_x$ to eigenvalue $-j$, i.e.
\begin{equation}
 (\cos \theta \, J_z  - \sin \theta \,  J_x) |\theta\rangle = - j |\theta\rangle
\end{equation}

Expectation values of spin operators in the coherent state are  given by
\begin{equation}
\begin{split}
\langle \theta | J_x |\theta \rangle &= j \sin\theta \; , \\
\langle \theta | J_z | \theta \rangle &= -j \cos\theta \; .
\end{split}
\end{equation}

The overlap between two spin coherent states is given by 
\begin{align}
\langle \theta | \chi \rangle = \cos^{2j}\frac{\theta-\chi}{2} \;.
\end{align}
For $j \to \infty$, the overlap is zero for $\theta \ne \chi$.

\section{Spin variance}\label{app:JVar}

The oscillator variance $\Delta_q$
is invariant under translations, which modify the bosonic operators through a linear shift $a \mapsto a + \alpha$.
For an analogous spin variance we require invariance under rotations,
which leads to a slightly more complicated definition.
We restrict ourselves to real spin states,
the generalization to arbitrary spin states is straightforward.

We define the spin variance $\Delta_J$
as the minimal variance of a rotated spin operator $J_\parallel = \cos \theta J_z + \sin \theta J_x$
that is obtained through variation of the rotation angle $\theta$.
Expansion of $J_\parallel^2$ shows that $\Delta_J$ is the minimum of a quadratic form,
and given by the smaller eigenvalue of the $2 \times 2$ matrix
\begin{equation}\label{app:JVarMat}
  \begin{pmatrix}
  \langle J_x^2 \rangle - \langle J_x \rangle^2 &
  \langle J_x J_z \rangle - \langle J_x \rangle \langle J_z \rangle 
   \\[1ex]
  \langle J_x J_z \rangle - \langle J_x \rangle \langle J_z \rangle  
  &   \langle J_z^2 \rangle - \langle J_z \rangle^2
  \end{pmatrix} \;.
\end{equation}
Note that $\langle J_x J_z \rangle = \langle J_z J_x \rangle$ for a real spin state.

For a (real) spin coherent state $|\theta\rangle$,
we have $\Delta_J  = 0$
since $|\theta\rangle$ is obtained from rotation of the $J_z$-eigenstate $|j,-j\rangle$.
Conversely, if the smaller eigenvalue $\Delta_J = 0$,
the state is an eigenstate of $J_\parallel$, i.e. it is a rotated eigenstate $|j,m\rangle$ of $J_z$ (the angle $\theta$ could be deduced from the eigenvectors).
It need however not be a spin coherent state, which would require $m=\pm j$.

For the Dicke model in the $j\to \infty$-limit, the spin state for $\kappa<1$ is invariant under the $J_x \mapsto - J_x$ symmetry. The off-diagonal elements in Eq.~\eqref{app:JVarMat} vanish, and $\Delta_J = \Delta_z$ since $\langle J_x^2\rangle$ is of order $j$.

\end{document}